\documentclass[aps,pra,superscriptaddress,notitlepage,floatfix,twocolumn,longbibliography]{revtex4-1}
\usepackage{amssymb, amsmath, amsfonts} 
\usepackage{graphicx}
\usepackage[hidelinks,hypertexnames=false]{hyperref}
\usepackage[all]{hypcap}
\usepackage{docmute}

\renewcommand{\Re}{\operatorname{Re}}
\renewcommand{\Im}{\operatorname{Im}}

\newcommand{\Dca}{\Delta_\text{ca}}
\newcommand{\Dpc}{\Delta_\text{pc}}
\newcommand{\starkScalar}{\alpha_0}
\newcommand{\starkVector}{\alpha_1}
\newcommand{\mechW}{\omega_\text{m}}
\newcommand{\larmorW}{\omega_\text{s}}
\newcommand{\cavityW}{\omega_\text{c}}
\newcommand{\probeW}{\omega_\text{p}}
\newcommand{\kp}{k_\text{p}}
\newcommand{\Zho}{Z_\text{HO}}
\newcommand{\zho}{z_\text{HO}}
\newcommand{\atomN}{N}
\newcommand{\gs}{g_\text{s}}
\newcommand{\gm}{g_\text{m}}
\newcommand{\gsm}{g_\text{sm}}
\newcommand{\nbar}{\bar{n}}
\newcommand{\Sx}{\hat{F}_x}

\newcommand{\Sz}{\hat{F}_z}
\newcommand{\Svec}{\hat{{\mathbf{F}}}}
\newcommand{\Stot}{F}
\newcommand{\dMS}{\delta}
\newcommand{\kTot}{\Omega}
\newcommand{\kMS}{\Omega_\text{opt}}
\newcommand{\signM}{\epsilon}

\newcommand{\Zm}{\hat{Z}_\text{m}}
\newcommand{\Zs}{\hat{Z}_\text{s}}
\newcommand{\Pm}{\hat{P}_\text{m}}
\newcommand{\Ps}{\hat{P}_\text{s}}

\newcommand{\tc}{t_\text{c}}
\newcommand{\gc}{g_\text{c}}

\newcommand{\Dsq}{\langle \hat{D}^2 \rangle_\text{cyc}}

\begin{document}

\title{Negative-Mass Instability of the Spin and Motion of an Atomic Gas Driven by Optical Cavity Backaction}

\author{Jonathan Kohler}
\email{jkohler@berkeley.edu}
\affiliation{Department of Physics, University of California, Berkeley, California 94720, USA}
\author{Justin A. Gerber}
\affiliation{Department of Physics, University of California, Berkeley, California 94720, USA}
\author{Emma Dowd}
\affiliation{Department of Physics, University of California, Berkeley, California 94720, USA}
\author{Dan M.\ Stamper-Kurn}
\email{dmsk@berkeley.edu}
\affiliation{Department of Physics, University of California, Berkeley, California 94720, USA}
\affiliation{Materials Sciences Division, Lawrence Berkeley National Laboratory, Berkeley, California  94720, USA}

\begin{abstract}
We realize a spin-orbit interaction between the collective spin precession and center-of-mass motion of a trapped ultracold atomic gas, mediated by spin- and position-dependent dispersive coupling to a driven optical cavity.
The collective spin, precessing near its highest-energy state in an applied magnetic field, can be approximated as a negative-mass harmonic oscillator.
When the Larmor precession and mechanical motion are nearly resonant, cavity mediated  coupling leads to a negative-mass instability, driving exponential growth of a correlated mode of the hybrid system.
We observe this growth imprinted on modulations of the cavity field and estimate the full covariance of the resulting two-mode state by observing its transient decay during subsequent free evolution.
\end{abstract}

\maketitle

The description of a harmonic oscillator with negative mass applies to collective excitations in diverse non-equilibrium systems, such as solid-state crystals \cite{Ashcroft1976}, plasmas \cite{Nielsen1959, Kolomenskii1961}, superfluids \cite{Tucker1999}, and cold atomic gases \cite{Eiermann2003,Khamehchi2017}.
The total Hamiltonian describing a negative-mass harmonic oscillator has the opposite sign of that of a positive-mass oscillator, resulting in an inverted energy spectrum, where an increased oscillation amplitude lowers the total energy.
When a negative-mass oscillator is coherently coupled to a positive-mass oscillator at nearly the same frequency, the two-oscillator system can undergo an instability, where the transfer of energy between them leads to unbounded growth of the amplitudes of both oscillators.
This negative-mass instability has been observed in the classical mechanics of trapped plasmas \cite{Postma1966} and ion traps \cite{Strasser2002}, and has been suggested to play a role in galactic structure \cite{Lovelace1978}.

Negative-mass oscillators also play an important role in quantum science.
Glauber proposed that a negative-mass oscillator coupled to a zero-temperature bath would, through the negative-mass instability, function as an ideal quantum amplifier \cite{Glauber1986}.
Joint measurement of resonant, but uncoupled, positive- and negative-mass oscillators allows for continuous measurement in a backaction-free subspace \cite{Hammerer2009,Tsang2010}, recently demonstrated \cite{Ockeloen-Korppi2016,Møller2017} as a method to circumvent standard quantum limits for position and force detection.
Weak coupling of such modes, below the instability threshold, has been proposed for generation of steady-state, two-mode entanglement\ \cite{Andersen2012}.

In this Letter, we report experimental realization of the negative-mass instability in a fully quantum, optodynamical system.
Following Glauber \cite{Glauber1986} and recent experiments \cite{Krauter2011,Kohler2017,Møller2017}, the collective spin of an atomic gas with magnetic moments polarized opposite an applied magnetic field can be approximated as a negative-mass oscillator.
The positive-mass oscillator is provided by the center-of-mass motion of the same trapped atomic gas, cooled initially near its ground state.
A single-mode optical cavity introduces a third quantum element, which couples to each oscillator through magneto-optical and optomechanical interactions, respectively.
This cavity field both mediates interactions between the two oscillators, leading to a collective spin-orbit coupling within the atomic gas, and facilitates continuous measurement of the hybrid system, with precision near the standard quantum limits \cite{Schreppler2014}.
We observe amplification of both oscillators by the negative-mass instability, which, similar to a non-degenerate parametric amplifier, induces strongly correlated excitations in both modes.  We estimate the covariance of the resulting two-mode state from measurements of its transient decay during subsequent free evolution.
The observed gain and correlation amplitude are described well by a linearized model of the hybrid optodynamical system.

\begin{figure}[tb!]
	\includegraphics[width=3.375in]{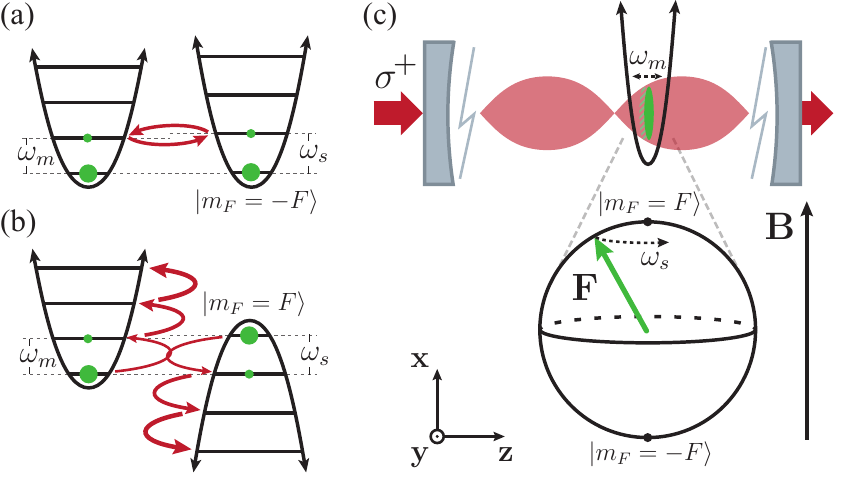}
	\caption{
		\label{fig:schematic}
		% red shows effective cavity mode volume (square root of intensity pattern...)
		(a-b) Energy levels of two nearly degenerate harmonic oscillators.
		(a) For positive-mass oscillators, coupling mediates exchange of excitations, conserving the total excitation number.
		(b) For positive- and negative-mass oscillators, the interaction resonantly drives pair creation, resulting in exponential growth of a correlated mode.
		(c) The center-of-mass motion of a harmonically confined, ultracold atomic ensemble (green), with trap frequency $\mechW$, represents the positive-mass oscillator.
		Larmor precession at frequency $\larmorW$ of the collective atomic spin near its highest-energy state, in an applied magnetic field $\mathbf{B} \propto \mathbf{x}$, approximates a negative-mass oscillator.		
		Position- and spin-dependent dispersive coupling to a circularly polarized mode of the optical cavity (red) mediates coherent interaction between the oscillators and facilitates continuous measurement of their dynamics.
	}
\end{figure}

To illustrate the negative-mass instability, consider a system of two harmonic oscillators, described by unitless bosonic operators $\hat{a}$ and $\hat{b}$, evolving at frequencies $\mechW = \omega_0 + \dMS/2$ and $\larmorW = \omega_0 - \dMS/2$, and let $\signM$ represent the sign of the second oscillator's mass.
If coupled by a spring of strength $\kTot$, the resulting dynamics are described by the interaction-picture Hamiltonian
\begin{align}
\label{eq:coupled-oscillators}
\mathcal{H}_I = \frac{\hbar \kTot}{2} \bigl(
\hat{a}^\dagger \hat{b} \, e^{i (\mechW - \signM \larmorW) t}
+ \hat{a}^\dagger \hat{b}^\dagger  e^{i (\mechW + \signM \larmorW) t}
+ h.c.
\bigr)
\text{.}
\end{align}
For nearly resonant oscillators, under the rotating-wave approximation, this interaction hybridizes their dynamics into coupled normal modes with eigenfrequencies $\omega_{\pm} = \omega_0 \pm \sqrt{ \dMS^2 + \signM \kTot^2 }/2$.  For positive masses ($\signM = +1$), the interaction results in a familiar avoided crossing in the energy spectrum and facilitates resonant exchange of excitations [Fig.~\ref{fig:schematic}a], conserving the total excitation number.

However, when the mass of the second oscillator is negative ($\signM = -1$), the pair-creation and pair-annihilation terms of the interaction are resonant [Fig.~\ref{fig:schematic}b], driving amplification of both oscillators.
For strong coupling ($| \kTot | > | \dMS |$), the normal-mode eigenvalues become complex, indicating the onset of the negative-mass instability.
In this condition, the oscillation frequencies of the normal modes, described by $\Re[\omega_{\pm}]$, synchronize, while the instability gain, described by $G_{\pm} = 2 \Im[\omega_{\pm}]$, indicates exponential amplification of one mode and damping of the other \cite{Gloppe2014}.
For resonant coupling ($\dMS = 0$), the amplified normal mode describes correlated motion of the two oscillators with a relative phase of $\pi/2$.
The resulting dynamics are similar to two-mode parametric amplification observed in driven optical four-wave mixing and down-conversion \cite{Lane1988}, which gives rise to the same equations of motion in a rotating frame defined by the optical pump \cite{Buchmann2015}.

%\section{System}

We experimentally realize the negative-mass instability using a gas of about 3000 $^{87}$Rb atoms, cooled to about 3\ $\mu$K by rf evaporation, and trapped in a single antinode of a standing-wave optical dipole trap (wavelength $842$ nm), resonant with a TEM$_{00}$ mode of a high-finesse, Fabry-P\'erot optical cavity\ \cite{Purdy2010}. For small displacements, the axial atomic motion is approximately harmonic, with trap frequency $\mechW$ controlled by the dipole trap intensity, defining a positive-mass, center-of-mass mode with unitless displacement $\Zm = \hat{a} + \hat{a}^\dagger$ defined in terms of bosonic phonon operators.

The ensemble is initially spin polarized in the $|f=2, m_f=2\rangle$ electronic ground state, yielding a total spin $\Stot \sim 6000$.
Applying a magnetic field along $\mathbf{x}$, transverse to the cavity axis, induces Larmor precession in the $\mathbf{y}$-$\mathbf{z}$ plane at frequency $\larmorW$.
For small, collective excitations of the total dimensionless spin $\Svec$ away from the magnetic field axis, the Larmor precession can be approximated as the motion of a harmonic oscillator, with unitless displacement defined as $\Zs = \sqrt{\Stot /2} \, \Sz = \hat{b} + \hat{b}^\dagger$ in terms of bosonic operators \cite{Holstein1940}.  The effective mass of this oscillator is negative (positive) for a spin precessing near its highest-energy (lowest-energy) state \cite{Kohler2017}.

The atomic ensemble is probed through its influence on another TEM$_{00}$ cavity mode, with half-linewidth $\kappa/2\pi = 1.82$\ MHz, which is detuned by $\Dca/2\pi=-42$\ GHz from the atomic $D_2$ transition, realizing an intensity- and spin-dependent dispersive coupling to circularly polarized light.
Positioning the trapped ensemble at the maximum intensity gradient of the probe field
 [Fig.~\ref{fig:schematic}c], its axial motion modulates the dispersive interaction, providing linear coupling to the center-of-mass displacement $\Zm$.
Optical coupling to the collective spin arises from the circular birefringence of the atomic ensemble\ \cite{Happer1967}.
For a cavity driven with circularly polarized light, this birefringence causes the dispersive coupling strength to depend linearly on $\Sz$, the projection of the total spin along the cavity axis, such that the cavity mode is coupled to one oscillating component of the transverse spin \cite{Brahms2010}.

Linearizing the collective dynamics for small excitations around an average cavity photon number $\nbar$, in a frame rotating at the optical probe frequency $\probeW$, results in an effective Hamiltonian
\footnote{See Supplemental Material at [SMURL], which includes Ref. \cite{Brahms2012}, for a derivation of the linearized Hamiltonian, and description of the instability fit procedure and matched filter analysis.}\nocite{Brahms2012}
\begin{multline}
\label{eq:coherent-hamiltonian}
\mathcal{H} =
\hbar \mechW \hat{a}^\dagger \hat{a}
+ \signM \hbar \larmorW \hat{b}^\dagger \hat{b}
- \hbar \Dpc \hat{c}^\dagger \hat{c}
\\
+ \hbar \sqrt{\nbar} ( \hat{c} + \hat{c}^\dagger )
\left[  \gm \Zm + \gs \Zs \right]
+ \hbar \nbar \gsm \Zm \Zs
\text{,}
\end{multline}
where $\Dpc=\probeW-\cavityW$ is the probe detuning from cavity resonance, $\hat{c}$  is the annihilation operator for photons in the cavity mode, and $\signM=-\operatorname{sgn} \langle \Sx \rangle$ is the sign of the spin oscillator's effective mass.
The coupling rates defined here are $\gs/2\pi = -18$~kHz, $\gm/2\pi = 26$~kHz, and $\gsm/2\pi = 120$~Hz for our system.

The coherent interactions between the three modes described by this Hamiltonian are more complicated than the model introduced in Eq.~\ref{eq:coupled-oscillators}.
However, the two-mode model can be recovered by adiabatic elimination of the cavity mode, in the unresolved sideband regime ($\kappa \gg \mechW, \larmorW$).
This results in optodynamical coupling between the collective motion and spin, with strength $\kMS = 4 \gs \gm \nbar \Dpc / ( \kappa^2 + \Dpc^2 )$ \cite{Spethmann2015}, in addition to independent optodynamical frequency shifts \cite{Sheard2004,Corbitt2006} and damping \cite{Arcizet2006,Gigan2006,Schliesser2006} of each oscillator.

The final term of Eq.~\ref{eq:coherent-hamiltonian} describes an additional, direct interaction between the motion and spin, which depends only on the mean photon number $\nbar$.
This `static' interaction arises from the spatial variation of the vector Stark shift and couples the motion and spin of each atom.  
Eq.~\ref{eq:coherent-hamiltonian} captures the projection of this interaction onto the collective modes, which, combined with the optodynamical coupling, results in a net spring strength $\kTot = \kMS + 2 \nbar \gsm$.  
In addition, there are residual incoherent dynamics, due to weak coupling between the spin and the thermal motion of each atom in the center-of-mass frame, which mediate a resonant, incoherent transfer of energy from the initially polarized spin into the mechanical bath, resulting in loss of spin polarization and anomalous diffusion of its precession \cite{Note1}.  

Owing to the spin- and position-dependent dispersive coupling in Eq.~\ref{eq:coherent-hamiltonian}, the probe field is sensitive to the joint displacement operator $\hat{D} = \gm \Zm + \gs \Zs$, imparting a state-dependent frequency shift to the effective cavity resonance.  
Through this shift, the oscillator dynamics are imprinted on phase and amplitude modulation of light transmitted through the cavity.  
These modulations are observed using an optical heterodyne detector, with total cavity photon detection efficiency $\varepsilon = 9\%$, from which a measurement record of $\hat{D}$ is recovered.

\begin{figure}[tb!]
\includegraphics[width=3.375in]{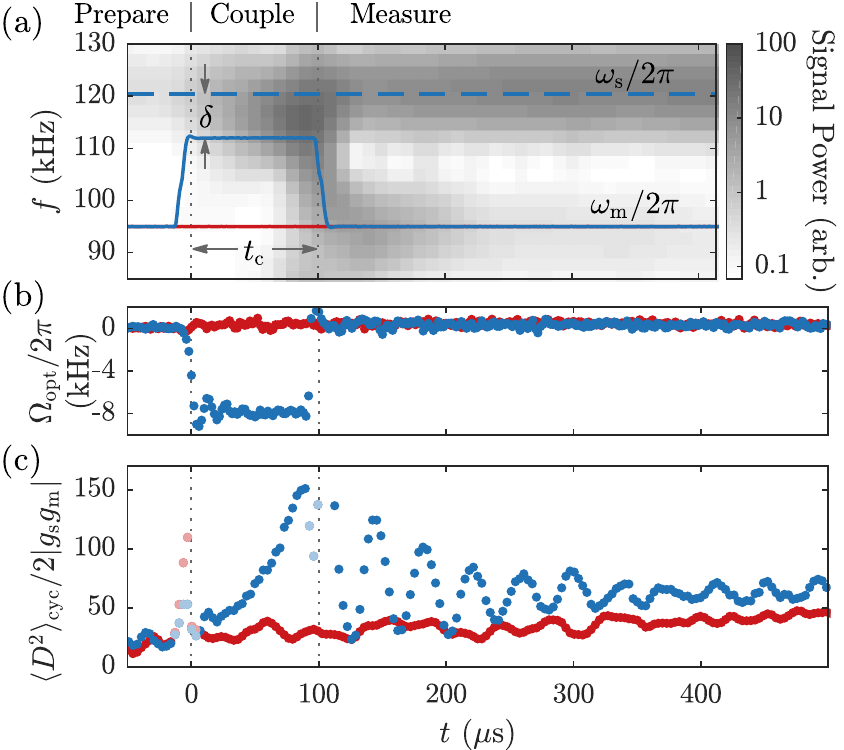}
\caption{
\label{fig:sequence}
Observation of negative-mass instability.
(a) Spectrogram of total optical modulation observed during the experimental sequence, averaged over $200$ iterations.  The spectrum shows components at the Larmor frequency, fixed at $\larmorW = 120$\ kHz (dashed line), and the mechanical frequency, initially at $\mechW=95$\ kHz then varied during coupling to achieve the desired detuning (solid line).
The collective spin shows negligible decay and the motion damps at rate $\Gamma_m/2\pi = 2$~kHz.
(b)
Optodynamical coupling strength $\kMS$, calculated from the measured $\Dpc$ and $\nbar$.
Experiments are performed with a coupling pulse with average $\nbar = 15$ and $\Dpc = 1.4$~MHz (blue) and without coupling (red).
(c) The mean squared joint displacement of both oscillators, captured in the cycle-averaged optical modulation power between $85$~kHz and $150$~kHz.
This signal reflects exponential amplification of both oscillators while coupled, followed by a stationary beat during the subsequent free evolution, revealing the transient decay of correlations created between the two modes.
Transients from changes in the optical probe and trap intensity perturb measurements near $t=0$ and $t=\tc$ (light points), which are excluded from analysis.
}
\end{figure}

%\section{Experiments}

We observe the negative-mass instability by applying a short optical coupling pulse to initially uncorrelated oscillators and measuring the subsequent free ringdown of the resulting state [Fig.~\ref{fig:sequence}].
During the initial preparation, the oscillator frequencies are well resolved to suppress interactions while the probe is stabilized on cavity resonance ($\Dpc = 0$) at a minimal intensity ($\nbar \approx 1$ \footnote{This probe intensity is chosen to minimize diffusion of the collective spin from coupling to thermal motion and accumulated measurement backaction, while providing sufficient signal-to-noise for the cavity-probe detuning feedback.}).
In the final stage of preparation, the mechanical frequency $\mechW$ is adiabatically ramped in 10 $\mu$s to achieve the desired detuning from the Larmor frequency $\dMS = \mechW - \larmorW$ [Fig.~\ref{fig:sequence}a].
The optical interaction is then quickly turned on for coupling time $\tc$ by increasing the probe intensity and stepping its detuning $\Dpc$ to achieve the desired coupling strength [Fig.~\ref{fig:sequence}b].
To observe the transient decay of the correlated two-mode state after the coupling pulse, the probe intensity is reduced ($\nbar \approx 4$),
 for improved measurement sensitivity, and the oscillator frequencies are resolved, by adiabatically ramping the optical trap back to its initial depth in 10 $\mu$s.

The coupled system evolves according to the projection of the oscillator's initial states onto the hybrid normal modes, where, under strong coupling, one mode is amplified and the other is damped.
Because both oscillators start near their ground states, without well-defined phases, the absolute phase of the amplified mode is random.
Therefore, each oscillator, observed independently, is driven into an effective thermal state with increased mean occupation, and the observed joint displacement $\langle \hat{D} \rangle$ averages to zero.
However, the growth of correlation between the oscillators results in motion with a fixed relative phase.

Both the amplification and correlation generated by the negative-mass instability are clearly captured in the cycle-averaged mean squared joint displacement
\begin{align*}
\label{eq:mean-squared-displacement}
\Dsq = \gm^2 \langle 2 \hat{a}^\dagger \hat{a} + 1 \rangle
+ \gs^2 \langle 2 \hat{b}^\dagger \hat{b} + 1\rangle
+ 4 \gm \gs \langle \Re[\hat{a} \, \hat{b}] \rangle
\text{.}
\end{align*}
Time evolution of this signal reveals the exponential growth of both oscillators during coupling, in addition to a stationary beat due to interference of the resulting two-mode correlations during subsequent free evolution [Fig.~\ref{fig:sequence}c].
This beat represents a self-heterodyne measurement arising from the product of the oscillator amplitudes, which evolves at their frequency difference, with initial amplitude and phase reflecting the magnitude and phase of correlation in the final state.

\begin{figure}[tb!]
\includegraphics[width=3.375in]{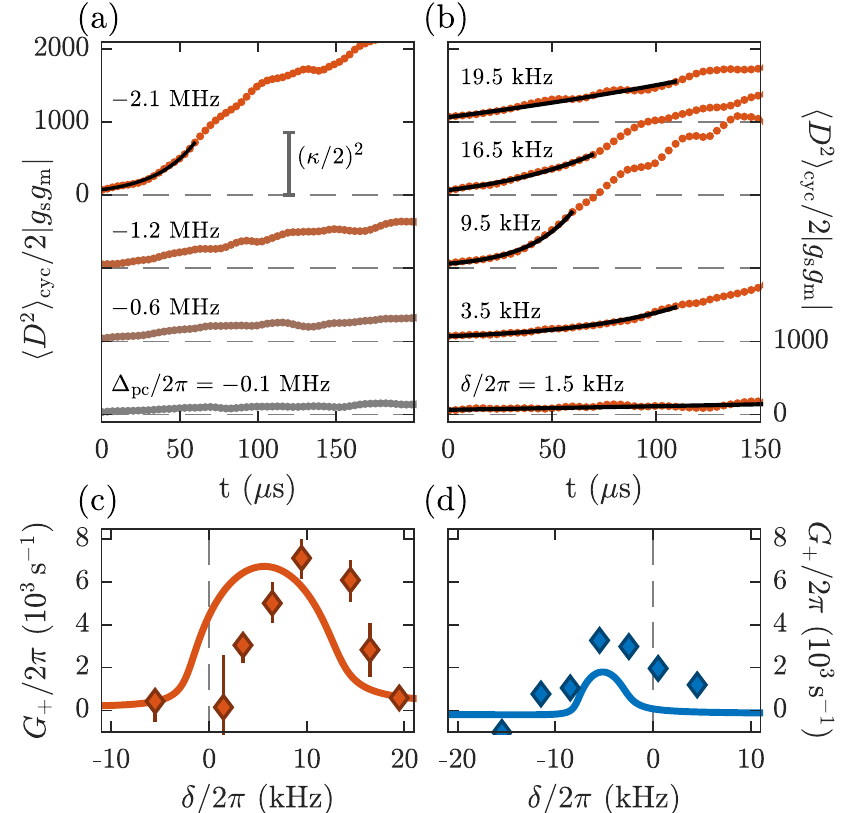}
\caption{
\label{fig:instability}
(a) Onset of instability, observed in the mean squared displacement, for increasing optical coupling strength, with $\nbar=10$ and oscillator detuning $\dMS/2\pi = 14$\ kHz.
Each trace (offset for clarity) is the average of around 30 repetitions.
Growth saturates due to the finite cavity linewidth (scale bar) and other non-linearities.
(b) Resonance of instability for varied $\dMS$, under strongest optical coupling in (a).
The instability gain $G_{+}$ is extracted by least-square fits (lines) to data at early times.
(c) 
$G_{+}$ (red diamonds) versus $\dMS$, compared to predicted steady-state gain (solid line).
The peak instability occurs at finite detuning, due to optodynamical shifts of each oscillator's frequency.
The larger frequency shift observed might be due to asymmetric transients, not reflected in the theoretical steady-state gain.
(d)
For an inverted probe detuning ($\Dpc = +2.0$\  MHz), the optodynamical coupling acts opposite the static coupling, resulting in reduced peak gain (blue diamonds).
Error bars in (c-d) represent combined 1-$\sigma$ statistical uncertainty from the fit and systematic error estimated from $\pm 10\%$ variations of the fit interval.
}
\end{figure}

%% Discussion of figure 3
The instability gain is measured in two ways---from growth of $\Dsq$ observed during coupling and from estimates of the resulting two-mode state after variable $\tc$.
Under strong coupling, both normal modes evolve at approximately the same frequency, such that $\Dsq$ predominately displays exponential growth at rate $G_{+}$, while incoherent dynamics driven by the thermal mechanical bath add diffusive growth to the observed signal.
For sufficiently strong optical coupling $\kTot$, near the optimal probe detuning $|\Dpc| = \kappa$,
the coherent interaction is dominant, resulting in the onset of instability observed in exponential growth of $\Dsq$  [Fig.~\ref{fig:instability}a].
The instability quickly drives the system into saturation, but for early times, the coherent growth is clearly reflected in the curvature of the measured signal.

We explore the instability's resonance by repeating these measurements over a range of $\dMS$ [Fig.~\ref{fig:instability}b].
The instability gain is extracted from the signal by a least-squares fit to a model describing coherent exponential amplification with additional diffusive noise, using independent rates to distinguish the coherent and incoherent dynamics \cite{Note1}.
The peak instability occurs at a non-zero detuning [Fig.~\ref{fig:instability}c], because independent optodynamical frequency shifts act on each oscillator in opposite directions, shifting them into resonance.
Inverting the sign of optical coupling $\kMS$, with an equal but opposite $\Dpc$, reveals the effect of the static coupling as an asymmetry in the observed gain [Fig.~\ref{fig:instability}d].

%\section{Correlations}

\begin{figure}[bt!]
\includegraphics[width=3.375in]{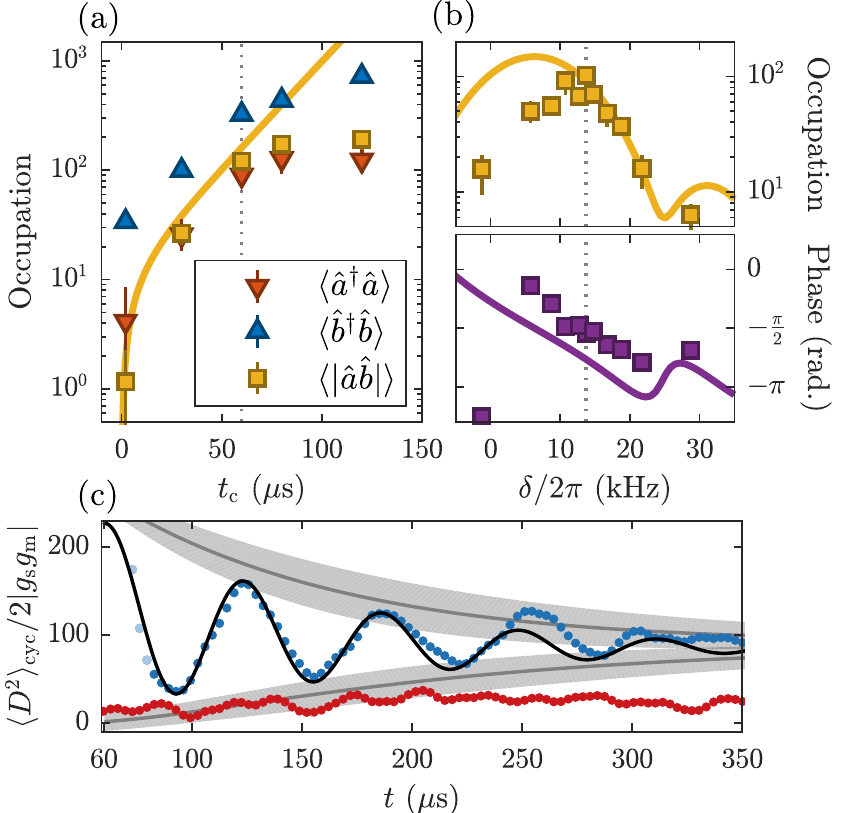}
\caption{
\label{fig:correlations}
Results of matched-filter analysis.
(a) Growth of the mechanical (red downward triangle), spin (blue upward triangle), and correlated (yellow squares) occupation as a function of $\tc$, with $\dMS/2\pi = 14$\ kHz and the same optical coupling as Fig.~\ref{fig:instability}c.
The observed correlated occupation agrees well with the predicted evolution of the measured initial state at the optimal detuning (yellow line).
Growth saturates near a mechanical occupation of 100, possibly due to mechanical non-linearity of the dipole trap.
(b) Correlated occupation and phase versus $\dMS$, after fixed $\tc=60\text{ }\mu$s.
Error bars in (a-b) indicate combined 1-$\sigma$ statistical uncertainty after 200 repetitions and estimated systematic error from the uncertainties of all filter parameters.
The anomalous frequency shift is similar to that seen in Fig.~\ref{fig:instability}c-d.
(c) Comparison between the beat observed in $\Dsq$ (blue points) and a simulated signal constructed from time evolution of the estimated two-mode state (black line) for one typical experimental setting.
Shaded regions show 1-$\sigma$ bounds for evolution of a maximally-correlated state with the same estimated individual oscillator occupations.  
Similar measurements performed without coupling (red points) show evolution of the initial state.
}
\end{figure}

The mean squared displacement, however, lacks spectral information distinguishing the occupation of each oscillator.
To estimate the final two-mode state $\hat{\mathbf{v}} = (\hat{a}^\dagger, \hat{b} )^\mathsf{T}$ from each experimental iteration, we apply linear matched filters directly to the free ringdown observed after the coupling pulse, extracting single-shot estimates for the amplitude and phase of each oscillator \cite{Palomaki2013,Spethmann2015}.
From an ensemble of measurements, we estimate the second-moment matrix $\mathrm{C} = \langle \hat{\mathbf{v}} \hat{\mathbf{v}}^\dagger \rangle$, correcting for correlated contributions from thermal noise, measurement backaction, and detector shot noise during the measurement interval \cite{Note1}.  

The diagonal components of the Hermitian matrix $\mathrm{C}$ capture the exponential growth of each oscillator's occupation for increasing $\tc$.
The off-diagonal component describes the amplitude and phase of correlation in the resulting state, which demonstrates the strong correlation of excitations added to both oscillators, providing an independent measure of $G_{+}$ unperturbed by the incoherent dynamics [Fig.~\ref{fig:correlations}a].
This amplitude and phase is measured across a range of $\dMS$, for a fixed $\tc=60$ $\mu$s, revealing the resonance of the correlation growth, with the expected correlation phase $\phi=-\pi/2$ at the optimal detuning [Fig.~\ref{fig:correlations}b].
We verify these matched-filter results by reconstructing the mean squared displacement from time evolution of the estimated covariance matrix $\mathrm{C}$ [Fig.~\ref{fig:correlations}c].

%\section{Conclusion}

In conclusion, we have demonstrated cavity-mediated coupling of the collective spin and motion of a trapped atomic ensemble.
For a high-energy polarized spin, this interaction results in a negative-mass instability, with dynamics analogous to a self-driven parametric amplifier.
We observed coherent amplification of a correlated mode by the instability, using time-resolved matched-filter analysis to estimate the covariance of the two-mode correlated state.
This instability could be applied as a coherent amplifier of an optomechanical state, facilitating enhanced measurement sensitivity, or to generate two-mode squeezed states, for use in entanglement enhanced metrology.
While, in our present system, any potential squeezing is obscured by incoherent coupling to thermal motion, this limitation could be avoided by using separate spin and mechanical oscillators, coupled only by cavity optodynamics.

\begin{acknowledgments}
This work was supported by the Air Force Office of Scientific Research.
J.~K. was supported by the U.S. Department of Defense through the National Defense Science and Engineering Graduate Fellowship program, and J.~G. and E.~D. by the National Science Foundation Graduate Fellowship.
\end{acknowledgments}

\bibliography{spinmech}

%\clearpage
\renewcommand{\bibliography}[1]{}

\title{Negative-mass instability of the spin and motion of an atomic gas driven by optical cavity backaction}

\maketitle

\renewcommand{\thesection}{S.\arabic{section}}
\renewcommand{\thesubsection}{\thesection.\arabic{subsection}}
\renewcommand{\thesubsubsection}{\thesubsection.\arabic{subsubsection}}
\renewcommand{\theequation}{S\arabic{equation}}
\renewcommand{\thefigure}{S\arabic{figure}}

\setcounter{section}{0}
\setcounter{subsection}{0}
\setcounter{subsubsection}{0}
\setcounter{equation}{0}
\setcounter{figure}{0}

% Fix references
\makeatletter
\renewcommand{\p@section}{}
\renewcommand{\p@subsection}{}
\renewcommand{\p@subsubsection}{}
\renewcommand{\p@equation}{}
\renewcommand{\p@figure}{}
\makeatother

\section*{Supplemental Material}

\section{Linearized Hamiltonian}
\label{sec:linear-hamiltonian}

Optomechanical coupling of center-of-mass atomic motion with an optical cavity has been previously described in Ref.~\cite{Purdy2010}, and the optodynamical interaction of collective spin precession with a cavity field is derived in Ref.~\cite{Brahms2010}.  Here we follow a similar derivation for simultaneous linear coupling of both the collective motion and spin to a single circularly-polarized cavity mode, which additionally yields a direct spin-orbit interaction and incoherent coupling of the spin to thermal atomic motion.

An ensemble of two-level atoms coupled to a cavity mode is well described by the Tavis-Cummings model.  In the dispersive limit, where the atom-cavity detuning $\Dca = \omega_c - \omega_a$ is large relative to the atomic linewidth, $\Dca \gg \Gamma_a$, the excited state can be adiabatically eliminated, resulting only in a dispersive shift to the cavity mode $\hat{c}$ and an ac Stark shift of the atomic ground state, captured by the Hamiltonian
\begin{align*}
\mathcal{H} = \hbar \cavityW \hat{c}^\dagger \hat{c} + \mathcal{H}_a
+ \hbar \sum_i^\atomN \hat{c}^\dagger \hat{c} \frac{ g(\hat{z}_i)^2 }{\Dca}
\text{,}
\end{align*}
where $\mathcal{H}_a$ contains the atomic energy, and $g(\hat{z}) = g_0 \sin( \kp \hat{z})$ is the position-dependent cavity-QED coupling strength, defined by the axial shape of the cavity mode and wavenumber $\kp$.

This model can be extended to describe multi-level atoms, such as $^{87}$Rb, assuming the cavity is sufficiently detuned from all excited state transitions.  For atoms in a ground-state hyperfine manifold $f$, then all $2f+1$ degenerate Zeeman sublevels can be coupled dispersively to the $f'=\left\{ f-1, f, f+1 \right\}$ excited-state hyperfine manifolds.  Summing the transition strengths to all the excited states yields a spin and polarization dependent dispersive coupling of the atoms to the cavity mode.  For a cavity driven with circularly polarized light and atom-cavity detuning $\Dca$ much greater than the hyperfine splitting, the effective Hamiltonian is
\begin{align*}
\mathcal{H} = \hbar \cavityW \hat{c}^\dagger \hat{c} + \mathcal{H}_a
+ \hbar \sum_i^\atomN  \hat{c}^\dagger \hat{c}
\left( \starkScalar + \starkVector \hat{ \mathbf{f} }_{i} \cdot \mathbf{z} \right) \frac{ g(\hat{z}_i)^2 }{\Dca} 
\text{,}
\end{align*}
where $\starkScalar=2/3$ and $\starkVector=1/6$ describe the relative strength of scalar and vector ac Stark shifts for $^{87}$Rb, and higher-order tensor interactions are negligible.

The atoms are trapped in a single potential well of a one-dimensional optical lattice along $\mathbf{z}$, formed by $842$ nm wavelength light resonating within the cavity.
For small displacements from the trap minimum at position $z_0$, this potential is approximately harmonic.
The axial motion of each atom at frequency $\mechW$ is described by bosonic phonon annihilation operators $\hat{a}_i$, and the radial motion is neglected.  
A large magnetic field $\boldsymbol{B}$ applied along $\mathbf{x}$, transverse to the cavity axis, induces Larmor precession of the total spin of each atom $\mathbf{f}_i$ at frequency $\larmorW= \mu_B |g_F \boldsymbol{B}|/\hbar$.  Including these dynamics, the total system Hamiltonian can be written
\begin{multline}
\mathcal{H} = \hbar \cavityW \hat{c}^\dagger \hat{c} 
+ \sum_i^\atomN \Bigl[
 \hbar \mechW \hat{a}_i^\dagger \hat{a}_i 
+ \hbar \larmorW \hat{f}_{x,i}
\\
+ \hbar \gc \hat{c}^\dagger \hat{c}
\left( \starkScalar + \starkVector \hat{f}_{z,i} \right)
\sin^2 { k_p (z_0 + \hat{z}_i) }
\Bigr]
\text{,}
\end{multline}
where $\hat{c}$ is the cavity photon annihilation operator, ${ \hat{f}_{i,j} = \hat{ \mathbf{f} }_i \cdot \mathbf{j} }$ is the component of atomic spin $i$ along axis $\mathbf{j}$, ${ \gc = g_0^2 / \Dca }$ is the dispersive cavity coupling strength, and ${ \hat{z}_i = \zho ( \hat{a}_i + \hat{a}_i^\dagger) }$ is the position of each atom, relative to the trap center, with harmonic-oscillator length ${ \zho = \sqrt{\hbar/2 m_a \mechW} }$.

Trapping the atoms at the maximum intensity gradient of the cavity mode ($\kp z_0 = 3 \pi/4$), and expanding to first order for small displacements around this position, 
\begin{multline}
\mathcal{H} = \hbar \left( \cavityW + \frac{\starkScalar}{2} \atomN \gc \right) \hat{c}^\dagger \hat{c} 
+ \hbar \mechW \sum_i^\atomN \hat{a}_i^\dagger \hat{a}_i 
+ \hbar \larmorW \Sx
\\
- \hbar \gc \hat{c}^\dagger \hat{c} 
\left[  \starkScalar k_p \sum_i^\atomN  \hat{z}_i - \frac{\starkVector}{2} \Sz + k_p \starkVector \sum_i^\atomN \hat{z}_i \hat{f}_{z,i} \right]
\text{.}
\end{multline}
where ${ \hat{F}_j = \sum_i \hat{f}_{j,i} }$ is the ensemble's total spin projection along axis $\mathbf{j}$.
In this expansion, the interaction causes a static cavity frequency shift $\cavityW' = \cavityW + \starkScalar \atomN \gc / 2$, a dispersive linear coupling of the collective atomic position and spin to the cavity mode, and a direct coupling of each atom's motion and spin.  This direct spin-orbit interaction arises because of the spatial variation of the vector ac Stark shift, which the atoms experience as an effective magnetic field gradient.

The atomic motion can be separated into a center-of-mass (CoM) mode $\hat{a} = \sum_i \hat{a}_i / \sqrt{\atomN}$ and the residual displacement of each atom relative to it, by making the substitution ${ \hat{a}_i \rightarrow \sqrt{1/\atomN} \hat{a} + \sqrt{(\atomN-1)/\atomN} \hat{a}_i }$.
The CoM position of the ensemble can then be described by the unitless displacement ${ \Zm = \hat{a} + \hat{a}^\dagger = \sum_i \hat{z}_i / (N \Zho) }$, with collective harmonic-oscillator length $\Zho = \zho / \sqrt{N}$, and conjugate momentum ${ \Pm = i ( \hat{a}^\dagger - \hat{a} ) }$.
The resulting Hamiltonian can be written as the sum of two parts, $\mathcal{H} = \mathcal{H}_c + \mathcal{H}_b$.   The `coherent' part $\mathcal{H}_c$ describes the dynamics and interactions of the collective modes 
\begin{multline}
\mathcal{H}_c
= \hbar \cavityW' \hat{c}^\dagger \hat{c} 
+ \hbar \mechW \hat{a}^\dagger \hat{a} 
+ \hbar \larmorW \Sx
- \hbar \gc \hat{c}^\dagger \hat{c} 
\\
\times
\Bigl[
\starkScalar \atomN \kp \Zho \Zm  
- \frac{\starkVector}{2} \Sz 
+ \starkVector \kp \Zho \Zm \Sz
\Bigr]
\text{,}
\end{multline}
and the mechanical `bath' part $\mathcal{H}_b$ contains the thermal energy of the residual atomic motion and a coupling between the spin and this effective mechanical bath,
\begin{multline}
\mathcal{H}_b
= \hbar \mechW \frac{N-1}{N} \sum_i^{\atomN} \hat{a}_i^\dagger \hat{a}_i 
\\
- \hbar \starkVector \kp \gc \sqrt{\frac{N-1}{N}} \hat{c}^\dagger \hat{c} \sum_i^{\atomN} \hat{z}_i \hat{f}_{z}^{(i)}
\text{.}
\end{multline}
The center-of-mass motion also equilibrates with this bath, likely due to collisions and trap inhomogeneity, which are not included in this model.

A rigorous treatment of the dynamics driven by this bath is beyond the scope of this work.  
However, considering a spin ensemble initially polarized in its highest-energy state and a steady-state coherent optical field, this interaction could facilitate a resonant, incoherent exchange of energy between the spin ensemble and the mechanical bath modes.
This model is consistent with an observed decay of spin polarization correlated with mechanical heating, as well as anomalous diffusion of the collective spin orientation [see Sec.~\ref{sec:incoherent}].

For small excitations of the collective spin away from either its highest- or lowest-energy state, its precession is approximated as the motion of a harmonic oscillator, described by bosonic operator $\hat{b}$, with unitless displacement ${ \Zs = \hat{b} + \hat{b}^\dagger }$ and conjugate momentum ${ \Ps = i(\hat{b}^\dagger - \hat{b}) }$, in terms of which, $\Sx = \operatorname{sgn} \langle \Sx \rangle (\Stot - \hat{b}^\dagger \hat{b})$ and $\Sz \approx \sqrt{\Stot / 2} \Zs$ \cite{Holstein1940}.
The dynamics of the cavity field are then linearized for small modulations about its average value, in a frame rotating at the probe frequency $\probeW$, by making the substitution $\hat{c} \rightarrow (\sqrt{\nbar} + \hat{c}) e^{-i \probeW t}$.  Keeping terms up to second order in the operators $\hat{a}$, $\hat{b}$, and $\hat{c}$, the coherent Hamiltonian is approximately
\begin{multline}
\label{eq:linearized-hamiltonian}
\mathcal{H}_c 
 =
- \hbar \Dpc \hat{c}^\dagger \hat{c} 
+ \hbar \mechW \hat{a}^\dagger \hat{a} 
+ \signM \hbar \larmorW  \hat{b}^\dagger \hat{b}
\\
+ \hbar \nbar \left[
\gm \Zm + \gs \Zs + \gsm \Zm \Zs
\right]
\\
+ \hbar \sqrt{\nbar} ( \hat{c} + \hat{c}^\dagger )
\left[
\gm \Zm + \gs \Zs
\right]
\text{,}
\end{multline}
where $\Dpc = \probeW - \cavityW'$ is the probe detuning from cavity resonance, $\signM = - \operatorname{sgn} \langle F_x \rangle$ indicates the sign of the spin's effective mass, negative (positive) for a spin near its highest-energy (lowest-energy) state, and defining coupling rates $\gm = -\starkScalar \atomN \gc \kp \Zho $, $\gs = \starkVector \gc \sqrt{\Stot / 8}  $, and $\gsm = -\starkVector \gc \kp \Zho  \sqrt{\Stot / 2}$.  

The second line of Eq. \ref{eq:linearized-hamiltonian} contains terms which depend only on the average probe intensity, and reflect constant, radiation pressure forces acting on each oscillator, in addition to a static coupling between the spin and mechanical modes.  Absorbing the constant forces into a negligible shift of the equilibrium position of each oscillator results in the linearized Hamiltonian quoted as Eq.\ 2 of the main text.

\section{Signatures of incoherent dynamics}
\label{sec:incoherent}

\begin{figure}[tb!]
	\includegraphics[width=3.375in]{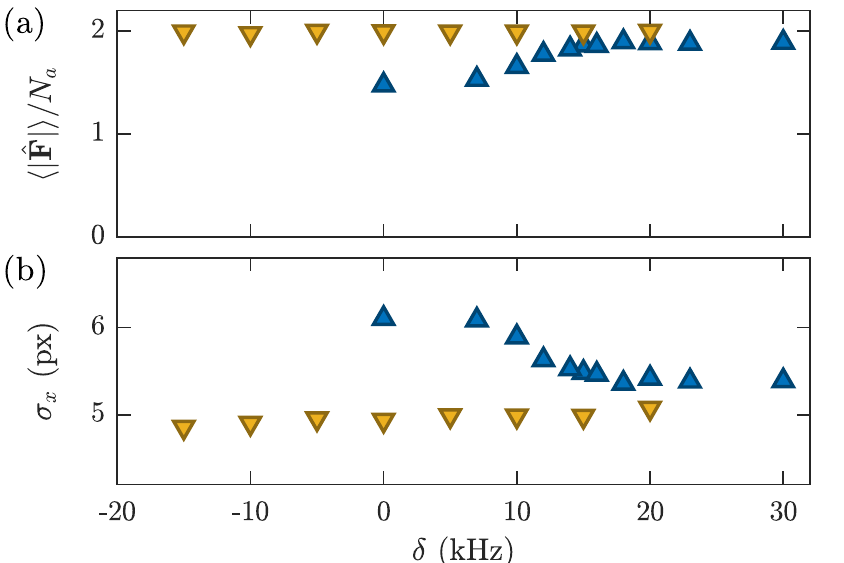}
	\includegraphics[width=3.375in]{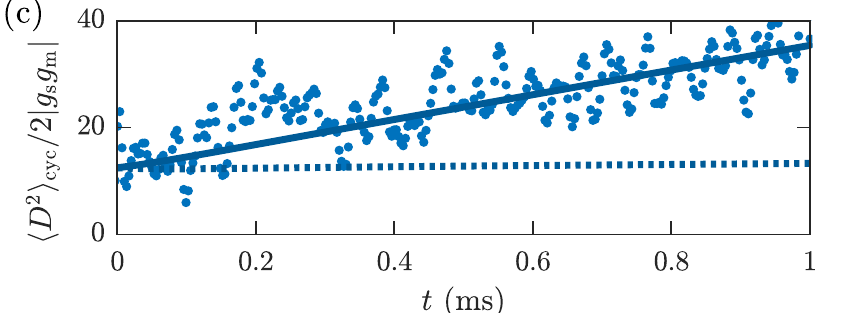}
	\caption{
		\label{fig:supp-incoherent}
		Signatures of incoherent dynamics.
		(a) Final spin polarization after coupling pulse (solid triangles).
		For an initially high-energy spin (blue, up triangles), the loss of polarization is largest near resonance with the mechanical oscillator.
		This loss is not observed for a low-energy spin, similarly coupled for $t_c = 300\text{ }\mu$s (yellow, down triangles)
		(b) Relative temperature of atomic ensemble, reflected in RMS width $\sigma_x$ of ensemble in absorption images after $0.5$ ms time-of-flight.
		The observed increase is correlated with loss of spin polarization, consistent with heating from incoherent energy exchange.
		(c) Linear growth of mean-squared joint displacement (blue dots), due to anomalous diffusion of spin under measurement conditions ($\nbar = 4$, $\Dpc = 0$, $\mechW / 2\pi = 95$~kHz, $\larmorW / 2\pi = 110$~kHz).
		The anomalous diffusion rate $B_s$ is determined by the difference between the expected measurement back-action rate (dotted line) and the observed linear growth rate (solid line).  
	}
\end{figure}

Coupling between the motion and spin of each atom facilitates an incoherent exchange of energy, resulting in a change of temperature for both the spin and mechanical modes of the ensemble.

The final spin temperature is indicated by the ensemble's polarization.
During each experimental sequence, the amplitude of transverse spin precession is observed in the heterodyne signal, but knowledge of the longitudinal component $\hat{F}_x$ is necessary to infer the final polarization.
Therefore, after each measurement sequence, we adiabatically rotate the magnetic field parallel to the cavity axis, such that the dispersive shift depends on the constant, longitudinal spin component.
The average cavity resonance frequency shift is given by $\Delta_c = \gc (\alpha_0 \atomN + \alpha_1 \hat{F}_x) /2$ (where $\hat{F}_x$ refers to the longitudinal spin component along the initial field orientation, prior to rotation).
This shift can be measured by sweeping the probe frequency across cavity resonance and fitting for the maximum intensity of transmitted light.  
We perform this swept measurement twice, inverting the spin projection with a Landau-Zener sweep in between, to distinguish the final atom number and average spin projection. 
The inferred spin polarization, corresponding to data in Fig.~4b, shows a loss of polarization for a negative-mass (high-energy) spin when the mechanical motion is nearly resonant [Fig.~\ref{fig:supp-incoherent}a].

The kinetic temperature of the gas is measured at the end of each experiment by absorption imaging after $0.5$\ ms time-of-flight expansion.  
We infer the relative temperature from the width of a 1-d Gaussian fit to an integrated density profile, transverse to the cavity axis \cite{Brahms2012}.  These measurements show an increased temperature correlated with loss of spin polarization, which is largest near $\delta = 0$, where the static coupling to the mechanical bath is resonant [Fig.~\ref{fig:supp-incoherent}b].
This condition is in contrast to the negative-mass instability, with resonance observed near $\delta=15$~kHz, primarily due to collective optodynamical spring shifts, which distinguish the oscillation frequencies of the collective modes.

The incoherent exchange of energy also drives diffusion of the collective spin precession.
During the measurement interval, with off-resonant oscillators ($\larmorW/ 2 \pi = 110$~kHz and $\mechW / 2\pi = 95$~kHz) probed on cavity resonance ($\nbar = 4$, $\Dpc=0$), we observe anomalous, linear growth of the mean-squared displacement, which greatly exceeds the expected diffusion from measurement backaction.
Fig.~\ref{fig:supp-incoherent}c shows the growth observed during measurement of the initial oscillator states, without a coupling pulse, primarily corresponding to an increasing incoherent spin occupation.
We attribute this diffusion to off-resonant coupling of the collective spin to the mechanical bath, and model it during the measurement interval as a phenomenological diffusion process, with diffusion rate $B_s$ calibrated from the observed excess growth.
This diffusion is also the dominant source of thermal excitations in the initial spin state, due to the minimal intensity of probe light necessary during the preparation stage to stabilize the probe frequency on cavity resonance.

\section{Equations of motion}
\label{sec:equation-of-motion}

Equations of motion for the bosonic operators describing all three modes can be derived from the Hamiltonian in Eq.~2, adding input and output terms to account for coupling to the mechanical bath and the optical vacuum.  It is also useful to define amplitude and phase quadratures of the optical field $\hat{c}_\text{AM} = (\hat{c} + \hat{c}^\dagger)/\sqrt{2}$ and $\hat{c}_\text{PM} = i (\hat{c} - \hat{c}^\dagger)/\sqrt{2}$, respectively.

The mechanical oscillator's equation of motion, assuming a phenomenological damping rate $\Gamma_m$ to a Markovian thermal bath $\hat{ \xi }_m$, is
\begin{multline}
\label{eq:eom-mech}
\dot{\hat{a}} = \left(- i \mechW - \frac{\Gamma_m}{2} \right) \hat{a}
- i \sqrt{2 \nbar} \gm \hat{c}_\text{AM}
- i \nbar \gsm \Zs
\\
+ \sqrt{\Gamma_m} \hat{\xi}_m
\text{,}
\end{multline}
The bath temperature is defined in terms of the oscillator's mean equilibrium occupation, initially $\bar{n}_m \sim 1$, with two-time correlation $\langle \hat{\xi}_m^\dagger(t) \hat{ \xi }_m(t') \rangle = \bar{n}_m \delta(t -t')$.
This temperature is estimated from the total scattered optical power observed in the mechanical sideband, calibrated by the sideband asymmetry observed in the heterodyne spectrum for a near-ground state oscillator \cite{Brahms2012}.
 
The spin oscillator is observed to have negligible loss for relevant timescales ($T_1 \gg 100$ ms, $T_2 > 5$ ms), other than the anomalous diffusion discussed in Sec.~\ref{sec:incoherent}.
During the measurement interval, this diffusion is included as a phenomenological Gaussian noise drive $\hat{ \xi }_s$, resulting in a spin equation of motion
 \begin{align}
 \label{eq:eom-spin}
 \dot{\hat{b}} & = i \signM \larmorW \hat{b}
 - i \sqrt{2 \nbar} \gs \hat{c}_\text{AM}
 - i \nbar \gsm \Zm
 + \sqrt{B_s} \xi_s
 \text{,}
\end{align}
with input noise correlation $\langle \xi_s(t) \xi_s(t') \rangle = \delta(t -t')$.  

Finally, the equation of motion for fluctuations of the optical field around the mean amplitude $\sqrt{\nbar}$ maintained by a coherent drive is
\begin{align}
\label{eq:eom-cav}
\dot{\hat{c}} & = (-\kappa + i \Delta) \hat{c}
- i \sqrt{\nbar} \Bigl[\gm \Zm + \gs \Zs \Bigl]
+ \sqrt{ 2 \kappa } \hat{\xi}_\text{in}
\text{,}
\end{align}
with cavity half-linewidth $\kappa$ and vacuum input noise $\hat{ \xi }_\text{in}$, with correlation $\langle \hat{\xi}(t) \hat{\xi}^\dagger(t') \rangle = \delta(t-t')$.

These equations form a complete linear system, which can be compactly represented by a matrix equation ${ \dot{ \mathbf{x} }  = \mathrm{M} \mathbf{x} + \mathbf{n} }$ in terms of a vector of the Hermitian quadratures of each operator
$ { \mathbf{x}  = \begin{pmatrix}
\Zm & \Pm & \Zs & \Ps & \hat{c}_\text{AM} & \hat{c}_\text{PM}
\end{pmatrix}^{T} } $
by defining the dynamical matrix $\mathrm{M}$ and noise-input vector $\mathbf{n}$ according to Eqns.\ \ref{eq:eom-mech}-\ref{eq:eom-cav}.
The eigenvectors of $\mathrm{M}$ describe the normal modes of the coupled system, with the corresponding eigenvalues becoming complex under strong coupling.  
The imaginary component of the eigenvalue corresponding to the amplified mode is the theoretical instability gain, displayed in Fig.~3c-d of the main text.  

Because the matrix $\mathrm{M}$ is invertible, solutions for the time evolution of products of the operators can efficiently be calculated by diagonalizing the equations of motion in the basis of normal modes.  Predictions for the final correlation, displayed in Fig.~4a-b, were calculated in this way, based on measurements of the uncorrelated initial states of the oscillators.

\section{Cavity output and detection}
\label{sec:cavity-output}
For an optodynamical system in the adiabatic regime ($\kappa \gg \omega_{s/m}$), the solution of the cavity field is approximately
\begin{align}
\label{eq:cavity-field-adiabatic}
\hat{c} & =
	\sqrt{ \frac{ \nbar } { \kappa^2 + \Dpc^2 } } \hat{ D } e^{- i \phi_q}
	+ \frac{\sqrt{  2 \kappa } } { \kappa - i \Delta } \hat{\xi}_\text{in}
\text{,}
\end{align}
in terms of the joint measurement operator $\hat{D}$ and defining quadrature angle $\tan \phi_q = \kappa / \Dpc$.
The amplitude and phase modulations of light transmitted from the cavity are detected by combining with a phase-coherent local oscillator (LO), offset by $10$\ MHz, with power $P_\text{LO} = 1$\ mW, and measuring the resulting heterodyne beatnote on a balanced photodiode, with total cavity photon detection efficiency $\varepsilon = 9\%$.

The optical power detected in the demodulated in-phase and quadrature-phase signal components is given by 
\begin{subequations}
\label{eq:quadratures}
\begin{align}
P_{I}(t) & = A \cos \phi_q \hat{ D }
+  \sqrt{ S_\text{SN} } \hat{\xi}_{+}
\text{ and }
\\
P_{Q}(t) & = A \sin \phi_q \hat{ D }
+  \sqrt{ S_\text{SN} } \hat{\xi}_{-}
\text{,}
\end{align}
\end{subequations}
respectively, with total detection gain ${ A = \sqrt{ \varepsilon S_\text{SN} \nbar \kappa / (\kappa^2 + \Dpc^2) } }$.
The second terms reflect measurement shotnoise (with power spectral density $S_\text{SN}= P_\text{LO} \hbar \omega_p$) arising from vacuum fluctuations of the probe field, described by quadrature operators $\hat{ \xi }_{\pm}$ with $\langle \hat{ \xi }_\pm(t) \hat{ \xi }_\pm(t') \rangle = \delta(t - t') /2$.

\section{Fit for instability gain}
\label{sec:instability-fit}
The cycle-averaged, mean-squared joint displacement $\Dsq$ is recovered from the demodulated quadratures.
First, both quadratures are bandpass filtered, with bandwidth $f_\text{BW}=65$\ kHz, to remove technical noise and reduce shotnoise.
The filtered signals are squared then added together, to capture modulations in both quadratures, independent of quadrature phase $\phi_q$.
Any residual second harmonics are removed from the squared signal by `cycle averaging' with a final low-pass filter at $120$\ kHz.
From Eqs.~\ref{eq:quadratures}, this yields
\begin{align}
\Dsq = 
\frac{ \langle P_I(t)^2 + P_Q(t)^2 \rangle_\text{cyc} } { A^2 } -  \frac{ S_\text{SN} f_\text{BW} } { A^2 }
\text{.}
\end{align}
The measurement gain $A$ is determined by independent measurements of the probe intensity and detuning, and the shotnoise power captured in the final term is calibrated by applying the same analysis to signals measured with an empty cavity.

For strong coupling, we expect the observed signal to be dominated by the amplified mode.  This mode is also subject to diffusion driven by the thermal mechanical bath, discussed in Sec.~\ref{sec:incoherent}.  If we assume for short times that this bath is Markovian, with phenomenological diffusion rate $B_s$, then considering the response of a single mode with intrinsic gain $G_{+}$ under the effect of diffusion suggests a fit function for the mean-squared signal of
\begin{align}
\label{eq:fit}
D_{\text{fit}}(t) = D_0 e^{G_{+} t} + \frac{B_s}{G_{+}} ( e^{G_{+} t} - 1)
\text{,}
\end{align}
with free parameters $D_0$, $G_{+}$, and $B_s$, where $D_0$ captures the mean-squared displacement of the initial state.

\begin{figure}[t!]
	\includegraphics[width=3.375in]{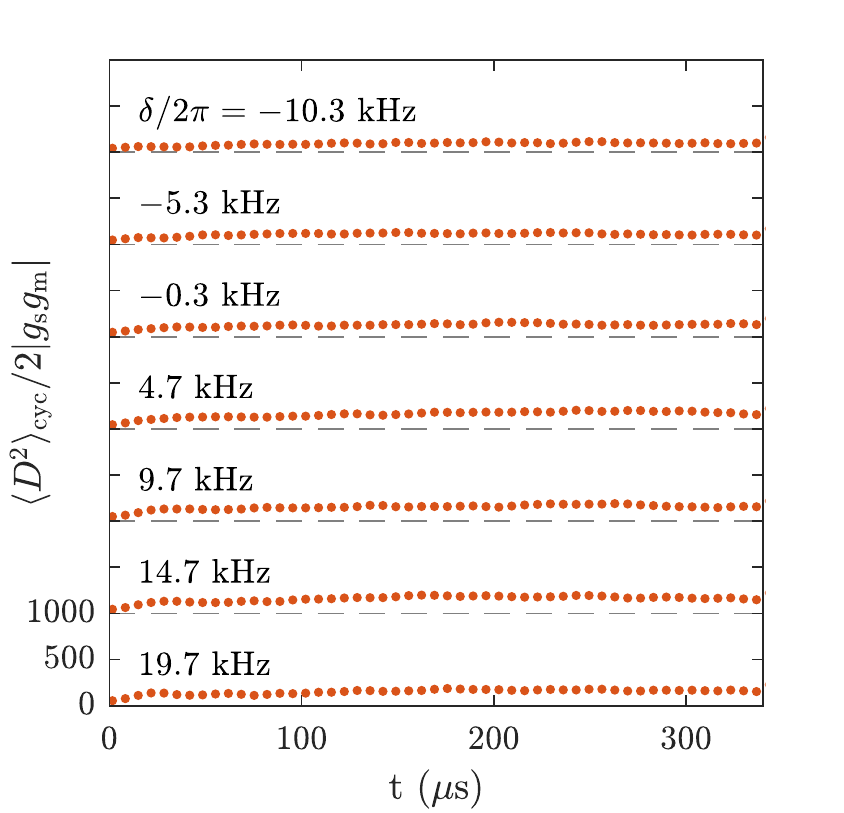}
	\caption{
		\label{fig:supp-posinstability}
		Absence of instability for two positive-mass oscillators, realized with spin polarized in lowest-energy state, coupled by the same optical spring.  Curves offset for clarity.
	}
\end{figure}

Because the instability gain quickly drives the system out of the linear regime, resulting in slower growth, the period of exponential growth observed in Fig.~3 is short.  
The choice of stop time for the fits is informed by monitoring the $\chi^2$ statistic.
For fit windows that are too short, the data does not constrain the fit well, resulting in large statistical uncertainty.
For fit windows that are too long, the data deviates significantly from an exponential model, increasing $\chi^2$. The reported fit stop times are chosen to fall within these limits and before the cavity shift becomes a significant fraction of the cavity linewidth, $\langle | \hat{D} | \rangle < \kappa/2$

To reduce the dimensionality of the fit, we take advantage of the identical preparation and probe conditions in Fig.~3b and simultaneously fit all the traces with a common initial amplitude $D_0$, but independent gains $G_{+}$ and diffusion rates $B_s$.
This procedure effectively constrains the initial amplitude for the fit of each individual trace, allowing the optimization to better distinguish between coherent and incoherent growth.
To estimate statistical uncertainties of the best estimates for these rates, we calculate the $\chi^2$ statistic across a 2-d region of parameter space around the optimal $G_{+}$ and $B_s$, identifying the  $\Delta \chi^2 = 1$ contour.  The range of each parameter defined by the contour determines the statistical error displayed in Fig.~3c-d, in addition to systematic errors estimated by varying the time interval of the fit by $\pm 10\%$.

For reference, we also performed a series of experiments similar to Fig.~3b, but for coupled positive-mass oscillators, with the spin prepared initially in its lowest energy state, verifying the absence of a negative-mass instability in this case [Fig.~\ref{fig:supp-posinstability}].

\section{Matched filter analysis}
\label{sec:matched-filter}

During measurement, with $\Dpc = 0$, the amplitude quadrature of the cavity field contains only the input vacuum fluctuations 
$\hat{c}_\text{AM} =  (\hat{\xi}_\text{in} + \hat{\xi}_\text{in}^\dagger) / \sqrt{ 2 \kappa }$ (for $\kappa \gg \omega_{m/s}$), which serve as the source of common backaction noise acting on each oscillator. In this condition, the equations of motion for each oscillator are independent and can be generalized (for $\hat{ \beta }_m = \hat{a}$, $\hat{ \beta }_s = \hat{b}$, and $i \in \{m,s\}$) as
\begin{align}
	\label{eq:general-eom}
	\dot{ \hat{\beta} }_i = \left( - i \signM_i \omega_i - \frac{ \Gamma_i }{2} \right) \hat{\beta}_i
	+ \hat{\eta}_i \text{,}
\end{align}
where the first term describes the coherent evolution, with $\signM_i$ the sign of mass for each oscillator, and $\hat{\eta}_i$ is the combined input noise, including the thermal noise driving each oscillator and common amplitude fluctuations of the cavity field.
The general solution to this equation, in terms of the initial state, is
\begin{align}
	\hat{\beta}_i(t) = \hat{\beta}_i(0) \rho_i(t)
	+ \int_0^t d\tau \hat{\eta}_i(\tau) \rho_i(t-\tau)
	\text{,}
\end{align}
defining an oscillator impulse response function $\rho_i(t) = \exp (-i \signM_i \omega_i t -\Gamma_i/2 t)$.

We recover estimates for the initial state of each oscillator at the beginning of the measurement interval from the outputs $q_i$ of a set of linear filters $m_i(t)$ applied to the recorded phase-modulation signal
\begin{align}
	\label{eq:filter-definition}
	q_i = \int_0^T P_Q(t) m_i(t) dt
	\text{.}
\end{align}
The filter functions are `matched' to the system dynamics by constructing them from the set of real and imaginary components of the complex oscillator response functions
\begin{align*}
r(t) = \Bigl(
\Re \left[  \rho_m(t) \right],
\Im \left[  \rho_m(t) \right],
\Re \left[  \rho_s(t) \right],
\Im \left[  \rho_s(t) \right] 
\Bigr)
\text{.}
\end{align*}
However, the signals in our experiment are complicated by shot-to-shot fluctuations of the oscillator frequencies.
Therefore, we define the set of matched filters in terms of components of the average response functions (indicated by an over bar),
\begin{align*}
m(t) = \Bigl(
\overline{ r_1(t) },
- \overline{ r_2(t) },
\overline{ r_3(t) },
- \overline{ r_4(t) }
\Bigr)
\text{.}
\end{align*}
Assuming the frequencies are normally distributed, with standard deviation $\sigma_i$ about mean $\omega_i$, the average complex response functions is ${ \overline{ \rho_i(t) } = \exp( -i \signM_i \omega_i t - \Gamma_i/2 t - \sigma_i^2 t^2/2 ) }$.
For the mechanical oscillator, the damping rate $\Gamma_m / 2\pi \sim 2$~kHz and inhomogeneous frequency distribution $\sigma_m / 2\pi \sim 500$~Hz are determined by fitting a Voigt profile to the observed power spectral density.  The spin has negligible intrinsic loss, so we assume $\Gamma_s = 0$ and determine $\sigma_s / 2\pi \sim 200$~Hz from the standard deviation of the Larmor frequency measured on each shot.

The filter results contain estimates for the state of each oscillator after coupling, in terms of the set of four Hermitian quadratures $ \hat{ X } = ( \Zm, \Pm, \Zs, \Ps )$. 
The expectation values of the quadrature operators are found by solving the linear set of equations
\begin{align}
\label{eq:filter-mean}
\langle q_i \rangle =A \sum_{j}^4 g_i \langle \hat{ X }_j  \rangle
\int_0^T dt \overline{ r_j(t) } m_i(t)
\text{,}
\end{align}
with $g_{1/2} = g_m$ and $g_{3/4} = g_s$ indexing the coupling strengths of the quadratures corresponding to each oscillator.
These equations describe how the oscillator quadratures are mixed in the filter outputs, due to slight non-orthogonality of the oscillators' response functions.

The second-moment matrix $\mathrm{C}$ can be expressed in terms of expectations of symmetrized products of quadrature operators $\langle \hat{X}_k \hat{X}_l + \hat{X}_l \hat{X}_k \rangle/2$.
These correlations are recovered by solving a similar set of equations for the second moments of filter outputs $\langle q_i q_j \rangle$, which also contain systematic biases from the non-zero variances of the input noise $\hat{ \eta }_i$ and measurement noise $\hat{ \xi }_{-}$ perturbing the observed signals.
In addition to increasing the variance of each filter's output, these noise sources also add to their covariances, again due to non-orthogonality of the oscillators' response functions.
We estimate the covariance due to measurement noise by applying the same set of filters to shotnoise-only signals, and numerically calculate the contribution of thermal noise and common measurement backaction from independent calibrations of all system and noise parameters.

The reliability of the final results depends on how well `matched' the filter functions are to the actual system evolution.
We estimate the systematic uncertainty of the measured correlations by sampling the distribution of results for variations of all the system and noise parameters used in the matched-filter analysis.
We generate sets of filters by randomly sampling each parameter from a normal distribution reflecting its calibrated mean and uncertainty and then compute the estimated occupations and correlations from the measured signals.
The systematic uncertainties plotted in Fig.~4a-b reflect the central $68$\% region of the resulting distributions, estimated from thousands of such randomly sampled filter sets.

\bibliography{spinmech}

\begin{thebibliography}{35}%
\makeatletter
\providecommand \@ifxundefined [1]{%
 \@ifx{#1\undefined}
}%
\providecommand \@ifnum [1]{%
 \ifnum #1\expandafter \@firstoftwo
 \else \expandafter \@secondoftwo
 \fi
}%
\providecommand \@ifx [1]{%
 \ifx #1\expandafter \@firstoftwo
 \else \expandafter \@secondoftwo
 \fi
}%
\providecommand \natexlab [1]{#1}%
\providecommand \enquote  [1]{``#1''}%
\providecommand \bibnamefont  [1]{#1}%
\providecommand \bibfnamefont [1]{#1}%
\providecommand \citenamefont [1]{#1}%
\providecommand \href@noop [0]{\@secondoftwo}%
\providecommand \href [0]{\begingroup \@sanitize@url \@href}%
\providecommand \@href[1]{\@@startlink{#1}\@@href}%
\providecommand \@@href[1]{\endgroup#1\@@endlink}%
\providecommand \@sanitize@url [0]{\catcode `\\12\catcode `\$12\catcode
  `\&12\catcode `\#12\catcode `\^12\catcode `\_12\catcode `\%12\relax}%
\providecommand \@@startlink[1]{}%
\providecommand \@@endlink[0]{}%
\providecommand \url  [0]{\begingroup\@sanitize@url \@url }%
\providecommand \@url [1]{\endgroup\@href {#1}{\urlprefix }}%
\providecommand \urlprefix  [0]{URL }%
\providecommand \Eprint [0]{\href }%
\providecommand \doibase [0]{http://dx.doi.org/}%
\providecommand \selectlanguage [0]{\@gobble}%
\providecommand \bibinfo  [0]{\@secondoftwo}%
\providecommand \bibfield  [0]{\@secondoftwo}%
\providecommand \translation [1]{[#1]}%
\providecommand \BibitemOpen [0]{}%
\providecommand \bibitemStop [0]{}%
\providecommand \bibitemNoStop [0]{.\EOS\space}%
\providecommand \EOS [0]{\spacefactor3000\relax}%
\providecommand \BibitemShut  [1]{\csname bibitem#1\endcsname}%
\let\auto@bib@innerbib\@empty
%</preamble>
\bibitem [{\citenamefont {Ashcroft}\ and\ \citenamefont
  {Mermin}(1976)}]{Ashcroft1976}%
  \BibitemOpen
  \bibfield  {author} {\bibinfo {author} {\bibfnamefont {Neil~W.}\ \bibnamefont
  {Ashcroft}}\ and\ \bibinfo {author} {\bibfnamefont {N.~David.}\ \bibnamefont
  {Mermin}},\ }\href@noop {} {\emph {\bibinfo {title} {{Solid state
  physics}}}}\ (\bibinfo  {publisher} {Saunders},\ \bibinfo {address} {New
  York},\ \bibinfo {year} {1976})\BibitemShut {NoStop}%
\bibitem [{\citenamefont {Nielsen}\ \emph {et~al.}(1959)\citenamefont
  {Nielsen}, \citenamefont {Sessler},\ and\ \citenamefont
  {Symon}}]{Nielsen1959}%
  \BibitemOpen
  \bibfield  {author} {\bibinfo {author} {\bibfnamefont {C.~E.}\ \bibnamefont
  {Nielsen}}, \bibinfo {author} {\bibfnamefont {A.~M.}\ \bibnamefont
  {Sessler}}, \ and\ \bibinfo {author} {\bibfnamefont {K.~R.}\ \bibnamefont
  {Symon}},\ }\bibfield  {title} {\enquote {\bibinfo {title} {{Longitudinal
  instabilities in intense relativistic beams}},}\ }in\ \href@noop {} {\emph
  {\bibinfo {booktitle} {Proc. 2nd Int. Conf. High-Energy Accel.
  Instruments}}},\ \bibinfo {editor} {edited by\ \bibinfo {editor}
  {\bibfnamefont {L.}~\bibnamefont {Kowarski}}}\ (\bibinfo  {publisher}
  {CERN},\ \bibinfo {address} {Geneva, Switzerland},\ \bibinfo {year} {1959})\
  pp.\ \bibinfo {pages} {239--252}\BibitemShut {NoStop}%
\bibitem [{\citenamefont {Kolomenskii}\ and\ \citenamefont
  {Lebedev}(1961)}]{Kolomenskii1961}%
  \BibitemOpen
  \bibfield  {author} {\bibinfo {author} {\bibfnamefont {A~A}\ \bibnamefont
  {Kolomenskii}}\ and\ \bibinfo {author} {\bibfnamefont {A~N}\ \bibnamefont
  {Lebedev}},\ }\bibfield  {title} {\enquote {\bibinfo {title} {{Beam stability
  in stacked orbits}},}\ }\href {\doibase 10.1088/0368-3281/3/1/112} {\bibfield
   {journal} {\bibinfo  {journal} {J. Nucl. Energy. Part C, Plasma Physics,
  Accel. Thermonucl. Res.}\ }\textbf {\bibinfo {volume} {3}},\ \bibinfo {pages}
  {44--45} (\bibinfo {year} {1961})}\BibitemShut {NoStop}%
\bibitem [{\citenamefont {Tucker}(1999)}]{Tucker1999}%
  \BibitemOpen
  \bibfield  {author} {\bibinfo {author} {\bibfnamefont {M.~A.}\ \bibnamefont
  {Tucker}},\ }\bibfield  {title} {\enquote {\bibinfo {title} {{Direct Evidence
  for R Rotons Having Antiparallel Momentum and Velocity}},}\ }\href {\doibase
  10.1126/science.283.5405.1150} {\bibfield  {journal} {\bibinfo  {journal}
  {Science (80-. ).}\ }\textbf {\bibinfo {volume} {283}},\ \bibinfo {pages}
  {1150--1152} (\bibinfo {year} {1999})}\BibitemShut {NoStop}%
\bibitem [{\citenamefont {Eiermann}\ \emph {et~al.}(2003)\citenamefont
  {Eiermann}, \citenamefont {Treutlein}, \citenamefont {Anker}, \citenamefont
  {Albiez}, \citenamefont {Taglieber}, \citenamefont {Marzlin},\ and\
  \citenamefont {Oberthaler}}]{Eiermann2003}%
  \BibitemOpen
  \bibfield  {author} {\bibinfo {author} {\bibfnamefont {B.}~\bibnamefont
  {Eiermann}}, \bibinfo {author} {\bibfnamefont {P.}~\bibnamefont {Treutlein}},
  \bibinfo {author} {\bibfnamefont {Th.}\ \bibnamefont {Anker}}, \bibinfo
  {author} {\bibfnamefont {M.}~\bibnamefont {Albiez}}, \bibinfo {author}
  {\bibfnamefont {M.}~\bibnamefont {Taglieber}}, \bibinfo {author}
  {\bibfnamefont {K.-P.}\ \bibnamefont {Marzlin}}, \ and\ \bibinfo {author}
  {\bibfnamefont {M.~K.}\ \bibnamefont {Oberthaler}},\ }\bibfield  {title}
  {\enquote {\bibinfo {title} {{Dispersion Management for Atomic Matter
  Waves}},}\ }\href {\doibase 10.1103/PhysRevLett.91.060402} {\bibfield
  {journal} {\bibinfo  {journal} {Phys. Rev. Lett.}\ }\textbf {\bibinfo
  {volume} {91}},\ \bibinfo {pages} {060402} (\bibinfo {year}
  {2003})}\BibitemShut {NoStop}%
\bibitem [{\citenamefont {Khamehchi}\ \emph {et~al.}(2017)\citenamefont
  {Khamehchi}, \citenamefont {Hossain}, \citenamefont {Mossman}, \citenamefont
  {Zhang}, \citenamefont {Busch}, \citenamefont {Forbes},\ and\ \citenamefont
  {Engels}}]{Khamehchi2017}%
  \BibitemOpen
  \bibfield  {author} {\bibinfo {author} {\bibfnamefont {M.~A.}\ \bibnamefont
  {Khamehchi}}, \bibinfo {author} {\bibfnamefont {Khalid}\ \bibnamefont
  {Hossain}}, \bibinfo {author} {\bibfnamefont {M.~E.}\ \bibnamefont
  {Mossman}}, \bibinfo {author} {\bibfnamefont {Yongping}\ \bibnamefont
  {Zhang}}, \bibinfo {author} {\bibfnamefont {Th.}\ \bibnamefont {Busch}},
  \bibinfo {author} {\bibfnamefont {Michael~McNeil}\ \bibnamefont {Forbes}}, \
  and\ \bibinfo {author} {\bibfnamefont {P.}~\bibnamefont {Engels}},\
  }\bibfield  {title} {\enquote {\bibinfo {title} {{Negative-Mass Hydrodynamics
  in a Spin-Orbit–coupled Bose-Einstein Condensate}},}\ }\href {\doibase
  10.1103/PhysRevLett.118.155301} {\bibfield  {journal} {\bibinfo  {journal}
  {Phys. Rev. Lett.}\ }\textbf {\bibinfo {volume} {118}},\ \bibinfo {pages}
  {155301} (\bibinfo {year} {2017})}\BibitemShut {NoStop}%
\bibitem [{\citenamefont {Postma}\ \emph {et~al.}(1966)\citenamefont {Postma},
  \citenamefont {Dunlap}, \citenamefont {Dory}, \citenamefont {Haste},\ and\
  \citenamefont {Young}}]{Postma1966}%
  \BibitemOpen
  \bibfield  {author} {\bibinfo {author} {\bibfnamefont {H.}~\bibnamefont
  {Postma}}, \bibinfo {author} {\bibfnamefont {J.~L.}\ \bibnamefont {Dunlap}},
  \bibinfo {author} {\bibfnamefont {R.~A.}\ \bibnamefont {Dory}}, \bibinfo
  {author} {\bibfnamefont {G.~R.}\ \bibnamefont {Haste}}, \ and\ \bibinfo
  {author} {\bibfnamefont {R.~A.}\ \bibnamefont {Young}},\ }\bibfield  {title}
  {\enquote {\bibinfo {title} {{Observation of Negative-Mass Instability in an
  Energetic Proton Plasma}},}\ }\href {\doibase 10.1103/PhysRevLett.16.265}
  {\bibfield  {journal} {\bibinfo  {journal} {Phys. Rev. Lett.}\ }\textbf
  {\bibinfo {volume} {16}},\ \bibinfo {pages} {265--268} (\bibinfo {year}
  {1966})}\BibitemShut {NoStop}%
\bibitem [{\citenamefont {Strasser}\ \emph {et~al.}(2002)\citenamefont
  {Strasser}, \citenamefont {Geyer}, \citenamefont {Pedersen}, \citenamefont
  {Heber}, \citenamefont {Goldberg}, \citenamefont {Amarant}, \citenamefont
  {Diner}, \citenamefont {Rudich}, \citenamefont {Sagi}, \citenamefont
  {Rappaport}, \citenamefont {Tannor},\ and\ \citenamefont
  {Zajfman}}]{Strasser2002}%
  \BibitemOpen
  \bibfield  {author} {\bibinfo {author} {\bibfnamefont {D.}~\bibnamefont
  {Strasser}}, \bibinfo {author} {\bibfnamefont {T.}~\bibnamefont {Geyer}},
  \bibinfo {author} {\bibfnamefont {H.~B.}\ \bibnamefont {Pedersen}}, \bibinfo
  {author} {\bibfnamefont {O.}~\bibnamefont {Heber}}, \bibinfo {author}
  {\bibfnamefont {S.}~\bibnamefont {Goldberg}}, \bibinfo {author}
  {\bibfnamefont {B.}~\bibnamefont {Amarant}}, \bibinfo {author} {\bibfnamefont
  {A.}~\bibnamefont {Diner}}, \bibinfo {author} {\bibfnamefont
  {Y.}~\bibnamefont {Rudich}}, \bibinfo {author} {\bibfnamefont
  {I.}~\bibnamefont {Sagi}}, \bibinfo {author} {\bibfnamefont {M.}~\bibnamefont
  {Rappaport}}, \bibinfo {author} {\bibfnamefont {D.~J.}\ \bibnamefont
  {Tannor}}, \ and\ \bibinfo {author} {\bibfnamefont {D.}~\bibnamefont
  {Zajfman}},\ }\bibfield  {title} {\enquote {\bibinfo {title} {{Negative Mass
  Instability for Interacting Particles in a 1D Box: Theory and
  Application}},}\ }\href {\doibase 10.1103/PhysRevLett.89.283204} {\bibfield
  {journal} {\bibinfo  {journal} {Phys. Rev. Lett.}\ }\textbf {\bibinfo
  {volume} {89}},\ \bibinfo {pages} {283204} (\bibinfo {year}
  {2002})}\BibitemShut {NoStop}%
\bibitem [{\citenamefont {Lovelace}\ and\ \citenamefont
  {Hohlfeld}(1978)}]{Lovelace1978}%
  \BibitemOpen
  \bibfield  {author} {\bibinfo {author} {\bibfnamefont {R.~V.~E.}\
  \bibnamefont {Lovelace}}\ and\ \bibinfo {author} {\bibfnamefont {R.~G.}\
  \bibnamefont {Hohlfeld}},\ }\bibfield  {title} {\enquote {\bibinfo {title}
  {{Negative mass instability of flat galaxies}},}\ }\href {\doibase
  10.1086/156004} {\bibfield  {journal} {\bibinfo  {journal} {Astrophys. J.}\
  }\textbf {\bibinfo {volume} {221}},\ \bibinfo {pages} {51} (\bibinfo {year}
  {1978})}\BibitemShut {NoStop}%
\bibitem [{\citenamefont {Glauber}(1986)}]{Glauber1986}%
  \BibitemOpen
  \bibfield  {author} {\bibinfo {author} {\bibfnamefont {Roy~J.}\ \bibnamefont
  {Glauber}},\ }\bibfield  {title} {\enquote {\bibinfo {title} {{Amplifiers,
  Attenuators, and Schr{\"{o}}dinger's Cat}},}\ }\href {\doibase
  10.1111/j.1749-6632.1986.tb12437.x} {\bibfield  {journal} {\bibinfo
  {journal} {Ann. N. Y. Acad. Sci.}\ }\textbf {\bibinfo {volume} {480}},\
  \bibinfo {pages} {336--372} (\bibinfo {year} {1986})}\BibitemShut {NoStop}%
\bibitem [{\citenamefont {Hammerer}\ \emph {et~al.}(2009)\citenamefont
  {Hammerer}, \citenamefont {Aspelmeyer}, \citenamefont {Polzik},\ and\
  \citenamefont {Zoller}}]{Hammerer2009}%
  \BibitemOpen
  \bibfield  {author} {\bibinfo {author} {\bibfnamefont {Klemens}\ \bibnamefont
  {Hammerer}}, \bibinfo {author} {\bibfnamefont {Markus}\ \bibnamefont
  {Aspelmeyer}}, \bibinfo {author} {\bibfnamefont {Eugene~S.}\ \bibnamefont
  {Polzik}}, \ and\ \bibinfo {author} {\bibfnamefont {Peter}\ \bibnamefont
  {Zoller}},\ }\bibfield  {title} {\enquote {\bibinfo {title} {{Establishing
  Einstein-Poldosky-Rosen Channels between Nanomechanics and Atomic
  Ensembles}},}\ }\href {\doibase 10.1103/PhysRevLett.102.020501} {\bibfield
  {journal} {\bibinfo  {journal} {Phys. Rev. Lett.}\ }\textbf {\bibinfo
  {volume} {102}},\ \bibinfo {pages} {020501} (\bibinfo {year}
  {2009})}\BibitemShut {NoStop}%
\bibitem [{\citenamefont {Tsang}\ and\ \citenamefont
  {Caves}(2010)}]{Tsang2010}%
  \BibitemOpen
  \bibfield  {author} {\bibinfo {author} {\bibfnamefont {Mankei}\ \bibnamefont
  {Tsang}}\ and\ \bibinfo {author} {\bibfnamefont {Carlton~M.}\ \bibnamefont
  {Caves}},\ }\bibfield  {title} {\enquote {\bibinfo {title} {{Coherent
  Quantum-Noise Cancellation for Optomechanical Sensors}},}\ }\href {\doibase
  10.1103/PhysRevLett.105.123601} {\bibfield  {journal} {\bibinfo  {journal}
  {Phys. Rev. Lett.}\ }\textbf {\bibinfo {volume} {105}},\ \bibinfo {pages}
  {123601} (\bibinfo {year} {2010})}\BibitemShut {NoStop}%
\bibitem [{\citenamefont {Ockeloen-Korppi}\ \emph {et~al.}(2016)\citenamefont
  {Ockeloen-Korppi}, \citenamefont {Damsk{\"{a}}gg}, \citenamefont
  {Pirkkalainen}, \citenamefont {Clerk}, \citenamefont {Woolley},\ and\
  \citenamefont {Sillanp{\"{a}}{\"{a}}}}]{Ockeloen-Korppi2016}%
  \BibitemOpen
  \bibfield  {author} {\bibinfo {author} {\bibfnamefont {C.~F.}\ \bibnamefont
  {Ockeloen-Korppi}}, \bibinfo {author} {\bibfnamefont {E.}~\bibnamefont
  {Damsk{\"{a}}gg}}, \bibinfo {author} {\bibfnamefont {J.-M.}\ \bibnamefont
  {Pirkkalainen}}, \bibinfo {author} {\bibfnamefont {A.~A.}\ \bibnamefont
  {Clerk}}, \bibinfo {author} {\bibfnamefont {M.~J.}\ \bibnamefont {Woolley}},
  \ and\ \bibinfo {author} {\bibfnamefont {M.~A.}\ \bibnamefont
  {Sillanp{\"{a}}{\"{a}}}},\ }\bibfield  {title} {\enquote {\bibinfo {title}
  {{Quantum Backaction Evading Measurement of Collective Mechanical Modes}},}\
  }\href {\doibase 10.1103/PhysRevLett.117.140401} {\bibfield  {journal}
  {\bibinfo  {journal} {Phys. Rev. Lett.}\ }\textbf {\bibinfo {volume} {117}},\
  \bibinfo {pages} {140401} (\bibinfo {year} {2016})}\BibitemShut {NoStop}%
\bibitem [{\citenamefont {M{\o}ller}\ \emph {et~al.}(2017)\citenamefont
  {M{\o}ller}, \citenamefont {Thomas}, \citenamefont {Vasilakis}, \citenamefont
  {Zeuthen}, \citenamefont {Tsaturyan}, \citenamefont {Balabas}, \citenamefont
  {Jensen}, \citenamefont {Schliesser}, \citenamefont {Hammerer},\ and\
  \citenamefont {Polzik}}]{Møller2017}%
  \BibitemOpen
  \bibfield  {author} {\bibinfo {author} {\bibfnamefont {Christoffer~B.}\
  \bibnamefont {M{\o}ller}}, \bibinfo {author} {\bibfnamefont {Rodrigo~A.}\
  \bibnamefont {Thomas}}, \bibinfo {author} {\bibfnamefont {Georgios}\
  \bibnamefont {Vasilakis}}, \bibinfo {author} {\bibfnamefont {Emil}\
  \bibnamefont {Zeuthen}}, \bibinfo {author} {\bibfnamefont {Yeghishe}\
  \bibnamefont {Tsaturyan}}, \bibinfo {author} {\bibfnamefont {Mikhail}\
  \bibnamefont {Balabas}}, \bibinfo {author} {\bibfnamefont {Kasper}\
  \bibnamefont {Jensen}}, \bibinfo {author} {\bibfnamefont {Albert}\
  \bibnamefont {Schliesser}}, \bibinfo {author} {\bibfnamefont {Klemens}\
  \bibnamefont {Hammerer}}, \ and\ \bibinfo {author} {\bibfnamefont
  {Eugene~S.}\ \bibnamefont {Polzik}},\ }\bibfield  {title} {\enquote {\bibinfo
  {title} {{Quantum back-action-evading measurement of motion in a negative
  mass reference frame}},}\ }\href {\doibase 10.1038/nature22980} {\bibfield
  {journal} {\bibinfo  {journal} {Nature}\ }\textbf {\bibinfo {volume} {547}},\
  \bibinfo {pages} {191--195} (\bibinfo {year} {2017})}\BibitemShut {NoStop}%
\bibitem [{\citenamefont {Andersen}\ and\ \citenamefont
  {M{\o}lmer}(2012)}]{Andersen2012}%
  \BibitemOpen
  \bibfield  {author} {\bibinfo {author} {\bibfnamefont {Christian~Kraglund}\
  \bibnamefont {Andersen}}\ and\ \bibinfo {author} {\bibfnamefont {Klaus}\
  \bibnamefont {M{\o}lmer}},\ }\bibfield  {title} {\enquote {\bibinfo {title}
  {{Squeezing of collective excitations in spin ensembles}},}\ }\href {\doibase
  10.1103/PhysRevA.86.043831} {\bibfield  {journal} {\bibinfo  {journal} {Phys.
  Rev. A}\ }\textbf {\bibinfo {volume} {86}},\ \bibinfo {pages} {043831}
  (\bibinfo {year} {2012})}\BibitemShut {NoStop}%
\bibitem [{\citenamefont {Krauter}\ \emph {et~al.}(2011)\citenamefont
  {Krauter}, \citenamefont {Muschik}, \citenamefont {Jensen}, \citenamefont
  {Wasilewski}, \citenamefont {Petersen}, \citenamefont {Cirac},\ and\
  \citenamefont {Polzik}}]{Krauter2011}%
  \BibitemOpen
  \bibfield  {author} {\bibinfo {author} {\bibfnamefont {Hanna}\ \bibnamefont
  {Krauter}}, \bibinfo {author} {\bibfnamefont {Christine~A.}\ \bibnamefont
  {Muschik}}, \bibinfo {author} {\bibfnamefont {Kasper}\ \bibnamefont
  {Jensen}}, \bibinfo {author} {\bibfnamefont {Wojciech}\ \bibnamefont
  {Wasilewski}}, \bibinfo {author} {\bibfnamefont {Jonas~M.}\ \bibnamefont
  {Petersen}}, \bibinfo {author} {\bibfnamefont {J.~Ignacio}\ \bibnamefont
  {Cirac}}, \ and\ \bibinfo {author} {\bibfnamefont {Eugene~S.}\ \bibnamefont
  {Polzik}},\ }\bibfield  {title} {\enquote {\bibinfo {title} {{Entanglement
  Generated by Dissipation and Steady State Entanglement of Two Macroscopic
  Objects}},}\ }\href {\doibase 10.1103/PhysRevLett.107.080503} {\bibfield
  {journal} {\bibinfo  {journal} {Phys. Rev. Lett.}\ }\textbf {\bibinfo
  {volume} {107}},\ \bibinfo {pages} {080503} (\bibinfo {year}
  {2011})}\BibitemShut {NoStop}%
\bibitem [{\citenamefont {Kohler}\ \emph {et~al.}(2017)\citenamefont {Kohler},
  \citenamefont {Spethmann}, \citenamefont {Schreppler},\ and\ \citenamefont
  {Stamper-Kurn}}]{Kohler2017}%
  \BibitemOpen
  \bibfield  {author} {\bibinfo {author} {\bibfnamefont {Jonathan}\
  \bibnamefont {Kohler}}, \bibinfo {author} {\bibfnamefont {Nicolas}\
  \bibnamefont {Spethmann}}, \bibinfo {author} {\bibfnamefont {Sydney}\
  \bibnamefont {Schreppler}}, \ and\ \bibinfo {author} {\bibfnamefont {Dan~M.}\
  \bibnamefont {Stamper-Kurn}},\ }\bibfield  {title} {\enquote {\bibinfo
  {title} {{Cavity-Assisted Measurement and Coherent Control of Collective
  Atomic Spin Oscillators}},}\ }\href {\doibase 10.1103/PhysRevLett.118.063604}
  {\bibfield  {journal} {\bibinfo  {journal} {Phys. Rev. Lett.}\ }\textbf
  {\bibinfo {volume} {118}},\ \bibinfo {pages} {063604} (\bibinfo {year}
  {2017})}\BibitemShut {NoStop}%
\bibitem [{\citenamefont {Schreppler}\ \emph {et~al.}(2014)\citenamefont
  {Schreppler}, \citenamefont {Spethmann}, \citenamefont {Brahms},
  \citenamefont {Botter}, \citenamefont {Barrios},\ and\ \citenamefont
  {Stamper-Kurn}}]{Schreppler2014}%
  \BibitemOpen
  \bibfield  {author} {\bibinfo {author} {\bibfnamefont {Sydney}\ \bibnamefont
  {Schreppler}}, \bibinfo {author} {\bibfnamefont {Nicolas}\ \bibnamefont
  {Spethmann}}, \bibinfo {author} {\bibfnamefont {Nathan}\ \bibnamefont
  {Brahms}}, \bibinfo {author} {\bibfnamefont {Thierry}\ \bibnamefont
  {Botter}}, \bibinfo {author} {\bibfnamefont {Maryrose}\ \bibnamefont
  {Barrios}}, \ and\ \bibinfo {author} {\bibfnamefont {Dan~M.}\ \bibnamefont
  {Stamper-Kurn}},\ }\bibfield  {title} {\enquote {\bibinfo {title} {{Optically
  measuring force near the standard quantum limit}},}\ }\href {\doibase
  10.1126/science.1249850} {\bibfield  {journal} {\bibinfo  {journal} {Science
  (80-. ).}\ }\textbf {\bibinfo {volume} {344}},\ \bibinfo {pages} {1486--1489}
  (\bibinfo {year} {2014})}\BibitemShut {NoStop}%
\bibitem [{\citenamefont {Gloppe}\ \emph {et~al.}(2014)\citenamefont {Gloppe},
  \citenamefont {Verlot}, \citenamefont {Dupont-Ferrier}, \citenamefont
  {Siria}, \citenamefont {Poncharal}, \citenamefont {Bachelier}, \citenamefont
  {Vincent},\ and\ \citenamefont {Arcizet}}]{Gloppe2014}%
  \BibitemOpen
  \bibfield  {author} {\bibinfo {author} {\bibfnamefont {A.}~\bibnamefont
  {Gloppe}}, \bibinfo {author} {\bibfnamefont {P.}~\bibnamefont {Verlot}},
  \bibinfo {author} {\bibfnamefont {E.}~\bibnamefont {Dupont-Ferrier}},
  \bibinfo {author} {\bibfnamefont {A.}~\bibnamefont {Siria}}, \bibinfo
  {author} {\bibfnamefont {P.}~\bibnamefont {Poncharal}}, \bibinfo {author}
  {\bibfnamefont {G.}~\bibnamefont {Bachelier}}, \bibinfo {author}
  {\bibfnamefont {P.}~\bibnamefont {Vincent}}, \ and\ \bibinfo {author}
  {\bibfnamefont {O.}~\bibnamefont {Arcizet}},\ }\bibfield  {title} {\enquote
  {\bibinfo {title} {{Bidimensional nano-optomechanics and topological
  backaction in a non-conservative radiation force field}},}\ }\href {\doibase
  10.1038/nnano.2014.189} {\bibfield  {journal} {\bibinfo  {journal} {Nat.
  Nanotechnol.}\ }\textbf {\bibinfo {volume} {9}},\ \bibinfo {pages} {920--926}
  (\bibinfo {year} {2014})}\BibitemShut {NoStop}%
\bibitem [{\citenamefont {Lane}\ \emph {et~al.}(1988)\citenamefont {Lane},
  \citenamefont {Reid},\ and\ \citenamefont {Walls}}]{Lane1988}%
  \BibitemOpen
  \bibfield  {author} {\bibinfo {author} {\bibfnamefont {A.~S.}\ \bibnamefont
  {Lane}}, \bibinfo {author} {\bibfnamefont {M.~D.}\ \bibnamefont {Reid}}, \
  and\ \bibinfo {author} {\bibfnamefont {D.~F.}\ \bibnamefont {Walls}},\
  }\bibfield  {title} {\enquote {\bibinfo {title} {{Quantum analysis of
  intensity fluctuations in the nondegenerate parametric oscillator}},}\ }\href
  {\doibase 10.1103/PhysRevA.38.788} {\bibfield  {journal} {\bibinfo  {journal}
  {Phys. Rev. A}\ }\textbf {\bibinfo {volume} {38}},\ \bibinfo {pages}
  {788--799} (\bibinfo {year} {1988})}\BibitemShut {NoStop}%
\bibitem [{\citenamefont {Buchmann}\ and\ \citenamefont
  {Stamper-Kurn}(2015)}]{Buchmann2015}%
  \BibitemOpen
  \bibfield  {author} {\bibinfo {author} {\bibfnamefont {Lukas}\ \bibnamefont
  {Buchmann}}\ and\ \bibinfo {author} {\bibfnamefont {D.~M.}\ \bibnamefont
  {Stamper-Kurn}},\ }\bibfield  {title} {\enquote {\bibinfo {title}
  {{Nondegenerate multimode optomechanics}},}\ }\href {\doibase
  10.1103/PhysRevA.92.013851} {\bibfield  {journal} {\bibinfo  {journal} {Phys.
  Rev. A}\ }\textbf {\bibinfo {volume} {92}},\ \bibinfo {pages} {013851}
  (\bibinfo {year} {2015})}\BibitemShut {NoStop}%
\bibitem [{\citenamefont {Purdy}\ \emph {et~al.}(2010)\citenamefont {Purdy},
  \citenamefont {Brooks}, \citenamefont {Botter}, \citenamefont {Brahms},
  \citenamefont {Ma},\ and\ \citenamefont {Stamper-Kurn}}]{Purdy2010}%
  \BibitemOpen
  \bibfield  {author} {\bibinfo {author} {\bibfnamefont {Thomas~P.}\
  \bibnamefont {Purdy}}, \bibinfo {author} {\bibfnamefont {D.~W.~C.}\
  \bibnamefont {Brooks}}, \bibinfo {author} {\bibfnamefont {Thierry}\
  \bibnamefont {Botter}}, \bibinfo {author} {\bibfnamefont {Nathan}\
  \bibnamefont {Brahms}}, \bibinfo {author} {\bibfnamefont {Z.-Y.}\
  \bibnamefont {Ma}}, \ and\ \bibinfo {author} {\bibfnamefont {Dan~M.}\
  \bibnamefont {Stamper-Kurn}},\ }\bibfield  {title} {\enquote {\bibinfo
  {title} {{Tunable Cavity Optomechanics with Ultracold Atoms}},}\ }\href
  {\doibase 10.1103/PhysRevLett.105.133602} {\bibfield  {journal} {\bibinfo
  {journal} {Phys. Rev. Lett.}\ }\textbf {\bibinfo {volume} {105}},\ \bibinfo
  {pages} {133602} (\bibinfo {year} {2010})}\BibitemShut {NoStop}%
\bibitem [{\citenamefont {Holstein}\ and\ \citenamefont
  {Primakoff}(1940)}]{Holstein1940}%
  \BibitemOpen
  \bibfield  {author} {\bibinfo {author} {\bibfnamefont {T.}~\bibnamefont
  {Holstein}}\ and\ \bibinfo {author} {\bibfnamefont {H.}~\bibnamefont
  {Primakoff}},\ }\bibfield  {title} {\enquote {\bibinfo {title} {{Field
  Dependence of the Intrinsic Domain Magnetization of a Ferromagnet}},}\ }\href
  {\doibase 10.1103/PhysRev.58.1098} {\bibfield  {journal} {\bibinfo  {journal}
  {Phys. Rev.}\ }\textbf {\bibinfo {volume} {58}},\ \bibinfo {pages}
  {1098--1113} (\bibinfo {year} {1940})}\BibitemShut {NoStop}%
\bibitem [{\citenamefont {Happer}\ and\ \citenamefont
  {Mathur}(1967)}]{Happer1967}%
  \BibitemOpen
  \bibfield  {author} {\bibinfo {author} {\bibfnamefont {W.}~\bibnamefont
  {Happer}}\ and\ \bibinfo {author} {\bibfnamefont {B.~S.}\ \bibnamefont
  {Mathur}},\ }\bibfield  {title} {\enquote {\bibinfo {title} {{Off-Resonant
  Light as a Probe of Optically Pumped Alkali Vapors}},}\ }\href {\doibase
  10.1103/PhysRevLett.18.577} {\bibfield  {journal} {\bibinfo  {journal} {Phys.
  Rev. Lett.}\ }\textbf {\bibinfo {volume} {18}},\ \bibinfo {pages} {577--580}
  (\bibinfo {year} {1967})}\BibitemShut {NoStop}%
\bibitem [{\citenamefont {Brahms}\ and\ \citenamefont
  {Stamper-Kurn}(2010)}]{Brahms2010}%
  \BibitemOpen
  \bibfield  {author} {\bibinfo {author} {\bibfnamefont {Nathan}\ \bibnamefont
  {Brahms}}\ and\ \bibinfo {author} {\bibfnamefont {Dan~M.}\ \bibnamefont
  {Stamper-Kurn}},\ }\bibfield  {title} {\enquote {\bibinfo {title} {{Spin
  optodynamics analog of cavity optomechanics}},}\ }\href {\doibase
  10.1103/PhysRevA.82.041804} {\bibfield  {journal} {\bibinfo  {journal} {Phys.
  Rev. A}\ }\textbf {\bibinfo {volume} {82}},\ \bibinfo {pages} {041804}
  (\bibinfo {year} {2010})}\BibitemShut {NoStop}%
\bibitem [{Note1()}]{Note1}%
  \BibitemOpen
  \bibinfo {note} {See Supplemental Material at [SMURL], which includes Ref.
  \cite {Brahms2012}, for a derivation of the linearized Hamiltonian, and
  description of the instability fit procedure and matched filter
  analysis.}\BibitemShut {Stop}%
\bibitem [{\citenamefont {Brahms}\ \emph {et~al.}(2012)\citenamefont {Brahms},
  \citenamefont {Botter}, \citenamefont {Schreppler}, \citenamefont {Brooks},\
  and\ \citenamefont {Stamper-Kurn}}]{Brahms2012}%
  \BibitemOpen
  \bibfield  {author} {\bibinfo {author} {\bibfnamefont {Nathan}\ \bibnamefont
  {Brahms}}, \bibinfo {author} {\bibfnamefont {Thierry}\ \bibnamefont
  {Botter}}, \bibinfo {author} {\bibfnamefont {Sydney}\ \bibnamefont
  {Schreppler}}, \bibinfo {author} {\bibfnamefont {Daniel W.~C.}\ \bibnamefont
  {Brooks}}, \ and\ \bibinfo {author} {\bibfnamefont {Dan~M.}\ \bibnamefont
  {Stamper-Kurn}},\ }\bibfield  {title} {\enquote {\bibinfo {title} {{Optical
  Detection of the Quantization of Collective Atomic Motion}},}\ }\href
  {\doibase 10.1103/PhysRevLett.108.133601} {\bibfield  {journal} {\bibinfo
  {journal} {Phys. Rev. Lett.}\ }\textbf {\bibinfo {volume} {108}},\ \bibinfo
  {pages} {133601} (\bibinfo {year} {2012})}\BibitemShut {NoStop}%
\bibitem [{\citenamefont {Spethmann}\ \emph {et~al.}(2015)\citenamefont
  {Spethmann}, \citenamefont {Kohler}, \citenamefont {Schreppler},
  \citenamefont {Buchmann},\ and\ \citenamefont
  {Stamper-Kurn}}]{Spethmann2015}%
  \BibitemOpen
  \bibfield  {author} {\bibinfo {author} {\bibfnamefont {Nicolas}\ \bibnamefont
  {Spethmann}}, \bibinfo {author} {\bibfnamefont {Jonathan}\ \bibnamefont
  {Kohler}}, \bibinfo {author} {\bibfnamefont {Sydney}\ \bibnamefont
  {Schreppler}}, \bibinfo {author} {\bibfnamefont {Lukas}\ \bibnamefont
  {Buchmann}}, \ and\ \bibinfo {author} {\bibfnamefont {Dan~M.}\ \bibnamefont
  {Stamper-Kurn}},\ }\bibfield  {title} {\enquote {\bibinfo {title}
  {{Cavity-mediated coupling of mechanical oscillators limited by quantum
  back-action}},}\ }\href {\doibase 10.1038/nphys3515} {\bibfield  {journal}
  {\bibinfo  {journal} {Nat. Phys.}\ }\textbf {\bibinfo {volume} {12}},\
  \bibinfo {pages} {27--31} (\bibinfo {year} {2015})}\BibitemShut {NoStop}%
\bibitem [{\citenamefont {Sheard}\ \emph {et~al.}(2004)\citenamefont {Sheard},
  \citenamefont {Gray}, \citenamefont {Mow-Lowry}, \citenamefont {McClelland},\
  and\ \citenamefont {Whitcomb}}]{Sheard2004}%
  \BibitemOpen
  \bibfield  {author} {\bibinfo {author} {\bibfnamefont {Benjamin~S.}\
  \bibnamefont {Sheard}}, \bibinfo {author} {\bibfnamefont {Malcolm~B.}\
  \bibnamefont {Gray}}, \bibinfo {author} {\bibfnamefont {Conor~M.}\
  \bibnamefont {Mow-Lowry}}, \bibinfo {author} {\bibfnamefont {David~E.}\
  \bibnamefont {McClelland}}, \ and\ \bibinfo {author} {\bibfnamefont
  {Stanley~E.}\ \bibnamefont {Whitcomb}},\ }\bibfield  {title} {\enquote
  {\bibinfo {title} {{Observation and characterization of an optical
  spring}},}\ }\href {\doibase 10.1103/PhysRevA.69.051801} {\bibfield
  {journal} {\bibinfo  {journal} {Phys. Rev. A}\ }\textbf {\bibinfo {volume}
  {69}},\ \bibinfo {pages} {051801} (\bibinfo {year} {2004})}\BibitemShut
  {NoStop}%
\bibitem [{\citenamefont {Corbitt}\ \emph {et~al.}(2006)\citenamefont
  {Corbitt}, \citenamefont {Ottaway}, \citenamefont {Innerhofer}, \citenamefont
  {Pelc},\ and\ \citenamefont {Mavalvala}}]{Corbitt2006}%
  \BibitemOpen
  \bibfield  {author} {\bibinfo {author} {\bibfnamefont {Thomas}\ \bibnamefont
  {Corbitt}}, \bibinfo {author} {\bibfnamefont {David}\ \bibnamefont
  {Ottaway}}, \bibinfo {author} {\bibfnamefont {Edith}\ \bibnamefont
  {Innerhofer}}, \bibinfo {author} {\bibfnamefont {Jason}\ \bibnamefont
  {Pelc}}, \ and\ \bibinfo {author} {\bibfnamefont {Nergis}\ \bibnamefont
  {Mavalvala}},\ }\bibfield  {title} {\enquote {\bibinfo {title} {{Measurement
  of radiation-pressure-induced optomechanical dynamics in a suspended
  Fabry-Perot cavity}},}\ }\href {\doibase 10.1103/PhysRevA.74.021802}
  {\bibfield  {journal} {\bibinfo  {journal} {Phys. Rev. A}\ }\textbf {\bibinfo
  {volume} {74}},\ \bibinfo {pages} {021802} (\bibinfo {year}
  {2006})}\BibitemShut {NoStop}%
\bibitem [{\citenamefont {Arcizet}\ \emph {et~al.}(2006)\citenamefont
  {Arcizet}, \citenamefont {Cohadon}, \citenamefont {Briant}, \citenamefont
  {Pinard},\ and\ \citenamefont {Heidmann}}]{Arcizet2006}%
  \BibitemOpen
  \bibfield  {author} {\bibinfo {author} {\bibfnamefont {O.}~\bibnamefont
  {Arcizet}}, \bibinfo {author} {\bibfnamefont {P.-F.}\ \bibnamefont
  {Cohadon}}, \bibinfo {author} {\bibfnamefont {T.}~\bibnamefont {Briant}},
  \bibinfo {author} {\bibfnamefont {M.}~\bibnamefont {Pinard}}, \ and\ \bibinfo
  {author} {\bibfnamefont {A.}~\bibnamefont {Heidmann}},\ }\bibfield  {title}
  {\enquote {\bibinfo {title} {{Radiation-pressure cooling and optomechanical
  instability of a micromirror}},}\ }\href {\doibase 10.1038/nature05244}
  {\bibfield  {journal} {\bibinfo  {journal} {Nat.}\ }\textbf {\bibinfo
  {volume} {444}},\ \bibinfo {pages} {71--74} (\bibinfo {year}
  {2006})}\BibitemShut {NoStop}%
\bibitem [{\citenamefont {Gigan}\ \emph {et~al.}(2006)\citenamefont {Gigan},
  \citenamefont {B{\"{o}}hm}, \citenamefont {Paternostro}, \citenamefont
  {Blaser}, \citenamefont {Langer}, \citenamefont {Hertzberg}, \citenamefont
  {Schwab}, \citenamefont {B{\"{a}}uerle}, \citenamefont {Aspelmeyer},\ and\
  \citenamefont {Zeilinger}}]{Gigan2006}%
  \BibitemOpen
  \bibfield  {author} {\bibinfo {author} {\bibfnamefont {S.}~\bibnamefont
  {Gigan}}, \bibinfo {author} {\bibfnamefont {H.~R.}\ \bibnamefont
  {B{\"{o}}hm}}, \bibinfo {author} {\bibfnamefont {M.}~\bibnamefont
  {Paternostro}}, \bibinfo {author} {\bibfnamefont {F.}~\bibnamefont {Blaser}},
  \bibinfo {author} {\bibfnamefont {G.}~\bibnamefont {Langer}}, \bibinfo
  {author} {\bibfnamefont {J.~B.}\ \bibnamefont {Hertzberg}}, \bibinfo {author}
  {\bibfnamefont {Keith~C.}\ \bibnamefont {Schwab}}, \bibinfo {author}
  {\bibfnamefont {D.}~\bibnamefont {B{\"{a}}uerle}}, \bibinfo {author}
  {\bibfnamefont {Markus}\ \bibnamefont {Aspelmeyer}}, \ and\ \bibinfo {author}
  {\bibfnamefont {A.}~\bibnamefont {Zeilinger}},\ }\bibfield  {title} {\enquote
  {\bibinfo {title} {{Self-cooling of a micromirror by radiation pressure}},}\
  }\href {\doibase 10.1038/nature05273} {\bibfield  {journal} {\bibinfo
  {journal} {Nat.}\ }\textbf {\bibinfo {volume} {444}},\ \bibinfo {pages}
  {67--70} (\bibinfo {year} {2006})}\BibitemShut {NoStop}%
\bibitem [{\citenamefont {Schliesser}\ \emph {et~al.}(2006)\citenamefont
  {Schliesser}, \citenamefont {Del'Haye}, \citenamefont {Nooshi}, \citenamefont
  {Vahala},\ and\ \citenamefont {Kippenberg}}]{Schliesser2006}%
  \BibitemOpen
  \bibfield  {author} {\bibinfo {author} {\bibfnamefont {A.}~\bibnamefont
  {Schliesser}}, \bibinfo {author} {\bibfnamefont {P.}~\bibnamefont
  {Del'Haye}}, \bibinfo {author} {\bibfnamefont {N.}~\bibnamefont {Nooshi}},
  \bibinfo {author} {\bibfnamefont {K.~J.}\ \bibnamefont {Vahala}}, \ and\
  \bibinfo {author} {\bibfnamefont {T.~J.}\ \bibnamefont {Kippenberg}},\
  }\bibfield  {title} {\enquote {\bibinfo {title} {{Radiation Pressure Cooling
  of a Micromechanical Oscillator Using Dynamical Backaction}},}\ }\href
  {\doibase 10.1103/PhysRevLett.97.243905} {\bibfield  {journal} {\bibinfo
  {journal} {Phys. Rev. Lett.}\ }\textbf {\bibinfo {volume} {97}},\ \bibinfo
  {pages} {243905} (\bibinfo {year} {2006})}\BibitemShut {NoStop}%
\bibitem [{Note2()}]{Note2}%
  \BibitemOpen
  \bibinfo {note} {This probe intensity is chosen to minimize diffusion of the
  collective spin from coupling to thermal motion and accumulated measurement
  backaction, while providing sufficient signal-to-noise for the cavity-probe
  detuning feedback.}\BibitemShut {Stop}%
\bibitem [{\citenamefont {Palomaki}\ \emph {et~al.}(2013)\citenamefont
  {Palomaki}, \citenamefont {Teufel}, \citenamefont {Simmonds},\ and\
  \citenamefont {Lehnert}}]{Palomaki2013}%
  \BibitemOpen
  \bibfield  {author} {\bibinfo {author} {\bibfnamefont {T~A}\ \bibnamefont
  {Palomaki}}, \bibinfo {author} {\bibfnamefont {J.~D.}\ \bibnamefont
  {Teufel}}, \bibinfo {author} {\bibfnamefont {R.~W.}\ \bibnamefont
  {Simmonds}}, \ and\ \bibinfo {author} {\bibfnamefont {K.~W.}\ \bibnamefont
  {Lehnert}},\ }\bibfield  {title} {\enquote {\bibinfo {title} {{Entangling
  Mechanical Motion with Microwave Fields}},}\ }\href {\doibase
  10.1126/science.1244563} {\bibfield  {journal} {\bibinfo  {journal} {Science
  (80-. ).}\ }\textbf {\bibinfo {volume} {342}},\ \bibinfo {pages} {710--713}
  (\bibinfo {year} {2013})}\BibitemShut {NoStop}%
\end{thebibliography}%

\end{document}